\documentclass[preprint]{aastex}
\input epsf

\newcommand{\fe}{\mbox{$\langle{\rm Fe}\rangle$}}
\newcommand{\feh}{\mbox{$[{\rm Fe/H}]$}}
\newcommand{\alphafe}{\mbox{$[\alpha/{\rm Fe}]$}}
\newcommand{\mgb}{\mbox{Mg$b$}}
\newcommand{\mgtwo}{\mbox{Mg$_2$}}

\newcommand{\msun}{\mbox{M$_{\odot}$}}

\newcommand{\vi}{\mbox{$V\!-\!I$}}

\newcommand{\viz}{\mbox{$(V\!-\!I)_0$}}
\begin{document}
\title{Keck Spectroscopy of Globular Clusters in the Sombrero Galaxy
 \footnote{Based on data obtained at the W.M. Keck Observatory, which is 
 operated as a scientific partnership among the California Institute of 
 Technology, the University of California and the National Aeronautics and 
 Space Administration.}
}

\author{S{\o}ren S. Larsen and Jean P. Brodie
  \affil{UC Observatories / Lick Observatory, University of California,
         Santa Cruz, CA 95064, USA}
  \email{soeren@ucolick.org and brodie@ucolick.org}
\and
  Michael A. Beasley and Duncan A. Forbes 
  \affil{Astrophysics \& Supercomputing, Swinburne University, Hawthorn VIC
         3122, Australia}
  \email{mbeasley@mania.physics.swin.edu.au and dforbes@astro.swin.edu.au}
}

\begin{abstract}
  We analyze high signal-to-noise spectra for 14 globular clusters (GCs) in 
the Sombrero galaxy, NGC~4594 = M104, obtained with the LRIS spectrograph
on the Keck I telescope.  We find generally good 
agreement between spectroscopic metallicities and previous photometric 
metallicites based on \vi\ colors. Further, we use measurements of various 
Balmer line indices on co-added spectra of metal-poor and metal-rich GCs
to estimate ages for the GC subpopulations and find ages of 10--15 Gyr, 
with preferred ages around 11--12 Gyr and no detectable differences between 
the metal-rich and metal-poor subpopulations.  However, because of model and 
measurement uncertainties, age differences of a few (2--3) Gyrs cannot be 
ruled out. We also measure alpha to iron abundance ratios from Mg and TiO 
features and find enhancements on the order of $\alphafe\sim+0.4$, similar to 
the \alphafe\ abundance ratios observed in other old stellar populations.
Finally, we combine our sample with radial velocities for 34 GCs from 
\citet{bri97} and obtain virial and projected masses for the Sombrero of 
$M_{\rm VT} \, = \, (8.1\pm1.8)\times10^{11} \, \msun$ and 
$M_P \, = \, (5.3\pm1.0)\times10^{11} \, \msun$ within 17 kpc, respectively.
Using 12 clusters within the central 4.5 kpc the corresponding numbers
are $M_{\rm VT} \, = \, (2.1\pm1.1)\times10^{11} \, \msun$ and
$M_P \, = \, (2.0\pm1.3)\times10^{11} \, \msun$.
\end{abstract}


\section{Introduction}

  The Sombrero galaxy (M104 = NGC~4594) is well known for its dominating
bulge/halo component which contributes with about 80\% of the total 
luminosity \citep{ken88,bag98}. This makes the Sombrero an important
intermediary case between later-type spirals and elliptical/S0 galaxies,
and an ideal place to study the relation between spiral bulges and early-type 
galaxies. One interesting aspect of such studies is to compare the
globular cluster (GC) systems of early- and late-type galaxies. The Sombrero
has a very rich GC system with an estimated total of $\sim2000$ clusters
\citep{har84,bri92,for97a}, about an order of magnitude more than the
Milky Way and M31 \citep{har96,fus93}, but comparable to
many early-type galaxies of the same luminosity.  Using archive WFPC2 
images, \citet[hereafter Paper I]{lfb01} detected a clearly bimodal color
distribution for the Sombrero GCs, confirming hints by \citet{for97a}.
The number of red (metal-rich) and blue GCs were found to be roughly equal,
in contrast to M31 and the Milky Way where the metal-poor GCs constitute
about 2/3 -- 3/4 of the total population, but again similar to the situation
in early-type galaxies. Thus, we suggested in Paper I that the metal-rich
GCs in the Sombrero are associated with the bulge rather than with its
disk, in which case a much more modest number of metal-rich GCs would have 
been expected.  The WFPC2 data presented in Paper I also showed that the mean 
colors of the two GC populations in the Sombrero are very similar to those 
in early-type galaxies as well as the Milky Way, M31 and other spiral galaxies 
\citep{fbl01}, with inferred mean metallicities of 
[Fe/H]$\sim-1.40$ and $-0.54$. 

Although there may not be a one-to-one correspondence between star formation 
and formation of globular clusters, information about the age distribution 
of GCs would presumably provide valuable insight about major star formation 
episodes in their host galaxies.  An important, unresolved issue is whether 
age differences exist between the GC subpopulations in galaxies, and how 
large such age differences might be. This question is still controversial even 
in our Galaxy, where GC ages can be derived from color-magnitude diagrams 
\citep{svb96,buo98,ros99}, so it is not surprising that it has proven to be 
very difficult to obtain reliable information about ages of extragalactic 
GCs from integrated photometry and/or spectroscopy.  As shown by 
\citet{frei95}, blue horizontal-branch morphology can
have a significant impact on observed H$\beta$ indices, the most commonly
used age indicator. More recently,
\citet{schia02} have discussed a number of problems in the modeling of 
globular cluster spectra, and concluded that uncertainties in the luminosity 
function of the red and asymptotic giant branches, $\alpha$-element 
enhancement, horizontal branch 
morphology and other factors can affect spectroscopic age estimates by at 
least 3--4 Gyr. Observationally, useful constraints on typical GC ages 
require very accurate measurements of Balmer line indices, as the separation 
between e.g.\ the 10 Gyr and 11 Gyr isochrones is only about 0.1 \AA\ in
the equivalent width of H$\beta$.
Furthermore, at lower metallicities ($\feh\la-1$), recent models 
(Maraston 2001, priv.\ comm., Bruzual \& Charlot 2001, priv.\ comm.) suggest 
that the Balmer lines no longer depend on age in a simple way for ages 
greater than about 10 Gyr. 

  A more indirect way to constrain the formation history of stellar populations
in general, including GCs, is by studying
their abundance patterns. The Milky Way bulge and halo populations, as well
as many early-type galaxies, show an enhancement in the abundances of 
$\alpha$-elements relative to the Fe-group compared to the Sun
\citep[][and references therein]{van95}. The enhancement is typically
around $\alphafe \sim +0.3$ \citep{car96} and is also seen in the Sombrero
bulge stars \citep{pel99}.  The most commonly adopted explanation is that 
different elements are produced by stars with different
lifetimes, of which supernovae of type Ia are generally assumed to have
the longest lifetimes and make the strongest contribution to Fe. Thus,
populations that formed on time-scales too short for SN Ia to have made
any significant contribution to the abundance patterns show ``enhancement''
of other elements relative to Fe \citep[e.g.][]{tin79}.  Therefore, 
studies of abundance patterns may provide an alternative chronometer for the 
chemical evolutionary history of galaxies, although the timescale for 
development of SN Ia remains very uncertain.

  Because of its relatively small discance and rich GC system, the Sombrero 
galaxy is an attractive target for detailed studies of individual GCs. With 
8--10 m class telescopes, high-quality spectra can be obtained for many of the
brighter clusters, and by co-adding spectra of several GCs it is possible
to achieve sufficient S/N to bring the errors down to a level where model
uncertainties rather than observational errors dominate. This makes it
possible to test whether the models yield internally consistent results,
e.g.\ by comparing age estimates based on different Balmer lines (H$\beta$, 
H$\gamma$, H$\delta$). 

  Spectroscopically, the GC system of the Sombrero has been studied by
\citet{bri97} who obtained rather low S/N spectra for 34 confirmed GCs.
By co-adding the spectra they found a mean metallicity of 
[Fe/H] = $-0.70\pm0.3$, but the S/N was too low to constrain metallicities
of individual clusters. In this paper we present new spectra for GCs in
the Sombrero, selected from the Paper I sample. Of 17 GC candidates, 14 are
confirmed as GCs.  Combining our data with those of \citet{bri97}, the total 
number of confirmed GCs in the Sombrero is increased to 48. The best of 
our spectra have a S/N of about 50 per pixel, allowing us to accurately
measure the metallicities of individual clusters and compare with previous
photometric estimates. By co-adding the spectra of blue (metal-poor) and 
red (metal-rich) clusters, we also put constraints on the ages 
(Section \ref{sec:popsynt}) and abundance ratios (Section \ref{sec:alpha})
of the Sombrero GC subpopulations. Finally, we use all available GC
radial velocities in the Sombrero to constrain the mass of the Sombrero 
galaxy (Section \ref{sec:mass}).

\section{Data}

  The data were obtained with the LRIS spectrograph \citep{oke95} on the 
Keck I telescope during two observing runs on 2001, April 28 and May 19--20. 
We utilized the newly installed blue side on LRIS (``LRIS-B'') and a dichroic 
splitting at 5600\AA\ to simultaneously obtain blue and red spectra, covering 
the regions $\sim$3500--5500\AA\ and $\sim$5700--8500\AA, respectively. 
On the blue side a 400 l/mm grism was used, providing a dispersion of 
1.74\AA\ pix$^{-1}$ and a resolution of $\sim8$\AA . On the red side we used 
a 600 l/mm grating blazed at 5000\AA, with a dispersion of 1.28\AA\ pix$^{-1}$ 
and a resolution of $\sim6$\AA . In addition to the multislit spectra of 
Sombrero GC candidates, we also obtained long-slit spectra for a number
of radial velocity standard stars, flux standards, and Lick/IDS standard
stars from \citet{wor94}.

  Because of anticipated problems with ``ghosts'' in LRIS-B, we applied 
2\arcsec\ offsets to the telescope pointing for some exposures. Although 
these offsets were intended to be parallel to the slitlets, we found that 
they inevitably included some component perpendicular to the slits as well, 
resulting in greatly reduced signal from our science targets. Thus, we
quickly decided to abandon this procedure.  Another problem
was related to the input coordinates. For the April run we used coordinates
measured directly on the ``central'' and ``inner halo'' WFPC2 images 
from Paper I, using the METRIC task in the
STSDAS package. Unfortunately, there were systematic differences between
the two WFPC2 pointings which made it impossible to align all objects in
the slitlets at the same time. For the May run we used direct images
from the April run to alleviate this problem, but as a result of these
slitmask alignment problems, we achieved a significantly lower S/N for some 
of the objects than could have been possible under optimal circumstances.

  Following initial processing of the images (bias subtraction, flatfielding),
the spectra were extracted using standard tools in the
IRAF SPECRED package\footnote{IRAF is distributed by the National Optical
Astronomical Observatories, which are operated by the Association of
Universities for Research in Astronomy, Inc.~under contract with the National
Science Foundation}. As the spectra were obtained over several hours,
significant shifts in the wavelength scale due to mechanical flexure were 
found to be 
present between individual integrations, amounting to up to $\approx5$ \AA . 
Assuming that all spectra in a given exposure were subject to the same shifts, 
the shifts were determined by cross-correlating the spectra of the brightest
objects in each exposure with the spectra in a reference exposure, taken
immediately before the arc spectra. For the red spectra, the wavelength 
zero-point was further fine-tuned using the NaD night sky lines at 
$\lambda\lambda$ 5890 and 5896 \AA . Finally, the individual spectra 
of each object were co-added, using a sigma-clipping algorithm to reject 
cosmic ray events. The S/N at each pixel in the combined spectra was estimated 
from the variance of the combined pixels. The total integration time was 
$5\frac{1}{2}$ hours.

  Table~\ref{tab:clusters} lists all science objects for which spectra were
obtained. The object IDs refer to Paper I, where the coordinates are listed.
For convenience, we have also included the $V,I$ photometry from Paper I.
Note that the brightest cluster, H2--22, was saturated on the WFPC2 data
used in Paper I, making its \vi\ color unreliable. The radial velocities in
Table~\ref{tab:clusters} are generally measured on the red spectra, for 
which the wavelength calibration zero-point could be checked using the 
night sky lines. The errors on the radial velocities in 
Table~\ref{tab:clusters} are random errors, based on the uncertainties
in the location of the cross-correlation peak reported by the FXCOR task.
These do not take into account various systematic errors, mostly
due to uncertainties in wavelength calibration. We estimate that such
errors are on the order $\sim0.5$\AA, or $\sim25$ km/s.  No skylines were 
present in the blue spectra, and differences of 100--150 km/s between radial 
velocities measured on the blue and red spectra were not uncommon. Another
problem with the wavelength calibration in the blue stems from the fact
that the calibration lamps in LRIS only had a few (5--7) suitable lines. The 
best fitting 2nd order polynomial wavelength solutions in the blue generally
had a significant residual scatter (1--2\AA ), presumably indicating 
that higher-order terms ought to be included in the wavelength solution,
but the small number of calibration lines made it impossible to include
higher-order terms. 

  The radial velocity of the Sombrero galaxy is listed as 1024 km/s
in the RC3 catalogue \citep{devau91}. If we exclude the three objects
in Table~\ref{tab:clusters} with RV$<$600 km/s, we are left with
a sample of 14 globular clusters. These have a mean radial velocity
of 1019 km/s and a dispersion of 246 km/s, in excellent agreement
with the radial velocity of the Sombrero itself. Of the three objects with
radial velocities less than 600 km/s, two (H2-06 and H2-10) are very compact
and essentially unresolved on the HST images with effective radii $\la$ 0.3 pc
\citep{lfb01} and are most likely foreground stars. The third (H2-27) is quite 
well-resolved on the HST images 
and has an effective radius of 3.4 pc, typical for globular clusters.
It therefore seems likely that this object is indeed a globular cluster.
The S/N of this spectrum is very poor and the cross-correlation signal
was not very strong, so the radial velocity may be in error. In fact,
using the blue spectrum of this object, we get a radial velocity of 1069 km/s. 
Because of the uncertain radial velocity and poor S/N we exclude this
object from further analysis in this paper.

  Before spectrophotometric indices were measured, the spectra must be
corrected for radial velocity. Rather than blindly applying the radial 
velocities in Table~\ref{tab:clusters} to the blue spectra, we used a
number of easily 
identifiable features such as the H$\beta$, H$\gamma$, H$\delta$, Ca II 
H$+$K lines and the G-band to determine small zero-point shifts and 
scale corrections to the dispersion solutions. Thus, the wavelength
scale of the final spectra used for the spectrophotometric analysis was
typically accurate to better than 1\AA\ in the region of interest
(i.e. $\sim3700$--5400 \AA ).

\section{Measuring Lick/IDS indices on LRIS-B spectra}
\label{sec:stdcmp}

  In order to test if there might be any systematic differences between
indices measured on LRIS-B spectra and the standard Lick/IDS system,
we observed a number of stars from \citet{wor94} in longslit mode. 
In Fig.~\ref{fig:stdcmp} we compare our LRIS-B measurements
of all Lick/IDS indices that were covered by our blue spectra with the
standard values. Generally, our measurements agree well with the
standard values, although there are slight systematic offsets for some of
the narrower indices like Ca4227 and Ca4455.  This may be attributed to
the slightly higher resolution of LRIS-B compared to the Lick/IDS
spectrograph.  

  Unfortunately, the wavelength range covered in longslit mode by LRIS is 
slightly different from the one covered in multislit mode and does not 
extend beyond 5300\AA .  Therefore, several key indices (such as \mgtwo, 
Fe5270 and Fe5335) redward of this limit could not be compared. Furthermore,
since this part of the spectrum is not covered by the flux standards,
the flux calibration of the multislit spectra beyond 5300\AA\ also becomes 
uncertain.  This is particularly problematic because the blue reflection 
efficiency of the dichroic already begins to drop around $\sim$5300\AA .

\section{Constraining ages and metallicities}

\subsection{Metallicities}

  We derived metallicities for each GC using the calibration by 
\citet{bh90}, but adopting the modification to the error on the
combined metallicity estimate described in \citet{lb02}.  Metallicities for 
each of the individual calibrators, as well as the combined metallicity 
estimates are listed in Table~\ref{tab:mettab}. For comparison we have also 
included photometric metallicities based on the \vi\ colors in 
Table~\ref{tab:clusters} and the calibration in \citet{kis98}.  The 
spectroscopic and photometric metallicity estimates are compared in 
Fig.~\ref{fig:metcmp} and generally agree quite well. Three representative
globular cluster spectra with low, intermediate and high metallicities
are shown in Fig.~\ref{fig:spectra}.

\subsection{Comparison with population synthesis models}
\label{sec:popsynt}

  High S/N spectra have the potential to remove the age-metallicity
degeneracy inherent in photometric studies. At least in principle, ages
are well constrained by Balmer line strengths, while metallicity information
is provided by indices such as Fe5270, Fe5335 and \mgtwo\ \citep{w94}. 

  Here we compare our Sombrero spectra with two sets of recent population 
synthesis models, kindly provided by C.\ Maraston 
\citep[see][]{mt00} and R.\
Schiavon \citep{scia02p}. Like most of the current models, the 
Maraston models are based on the ``fitting'' functions by \citet{wor94}.
The Schiavon models are based on a new set of fitting functions
derived from the spectral library of \citet{jon99}, and also include
empirical corrections to the luminosity function of red giants
\citep{schia02}.  We use a preliminary version of the Schiavon
models where only the Balmer line indices are based on the new fitting 
functions, i.e.\
other indices such as \mgb, \mgtwo\ and the various Fe indices are still
based on the Worthey fitting functions. The Schiavon models also cover
a more limited age/metallicity range than the Maraston models, but the
comparison nevertheless provides some useful insight into the differences
between various available models.
As discussed in Sect~\ref{sec:stdcmp}, our instrumental 
system appears to be very similar to the Lick/IDS system so we have not 
applied any corrections to spectrophotometric indices measured on the 
LRIS-B spectra. 

  As mentioned earlier, some indices may be compromised by difficulties
with the calibration of the blue spectra redward of 5300\AA . The
\mgtwo\ index is particularly vulnerable to such problems because of
the wide separation between the continuum passbands, and offsets are
frequently found between the standard Lick/IDS indices and instrumental
systems \citep[e.g.][]{kun00}.  In 
Fig.~\ref{fig:mgb_mg2} we compare our measurements of \mgtwo\ with \mgb ,
which is slightly narrower but measures essentially the same feature.
However, while the continuum passbands of \mgtwo\ are located about
150\AA\ from the feature passband, those of \mgb\ are located adjacent
to it, making \mgb\ much less sensitive to errors in the continuum 
slope.  Generally, Fig.~\ref{fig:mgb_mg2} shows that the two Mg sensitive 
indices correlate well, but the slope of our observed \mgb --\mgtwo\ relation 
is shallower than the Maraston model predictions, which in turn agree well 
with \mgb --\mgtwo\ plots in e.g. \citet{trag98} and \citet{beas00}.  One
possible explanation for this discrepancy is that errors in the continuum 
level at the position of the red continuum passband cause us to measure too
low values for the \mgtwo\ index because of uncorrected reduced throughput 
by the dichroic. Another possibility is that peculiarities in the
abundance patterns affect the continuum passbands of \mgb\ and \mgtwo\ in
different ways \citep{tb95}.

  In Fig.~\ref{fig:popmod} we compare various index measurements with the
Maraston and Schiavon model predictions. The Schiavon models are tabulated
at 5, 7.9, 11.2 and 14.1 Gyr and at metallicities of \feh\ = $-0.7$, $-0.4$,
0.0 and $+0.2$. While the Maraston models are tabulated at 1 Gyr intervals
from 2 to 15 Gyrs, we have plotted only the 5, 8, 11 and 14 Gyr models,
for easy comparison with the Schiavon models. Note that the metallicity
divisions of the Maraston models are different from those of the Schiavon
models (\feh = $-2.25$, $-1.35$, $-0.33$, 0.00 and $+0.35$).  Panels
(a) and (b) show H$\beta$ vs.\ \fe\ and the \mgb\ index, where 
\fe = (Fe5270 + Fe5335)/2. The Maraston and Schiavon model grids are shown 
with solid lines and dashed lines, respectively.  In panels (c) and (d) we 
show H$\gamma_A$ and H$\delta_A$ vs.\ \fe, using the definitions in 
\citet{wo97}.  Because of the difficulties with the calibration 
of \mgtwo\ discussed earlier, we have chosen \mgb\ instead of \mgtwo\ even 
though the formal errors on \mgb\ are somewhat larger. The errors on the
indices were estimated directly from the S/N of the spectra.

  To first order, Fig.~\ref{fig:popmod} indicates high ages for most
clusters, although the error bars are generally still too large to put
strong constraints on the ages of individual clusters. In order to compare 
the mean properties of metal-rich and metal-poor GCs, we therefore co-added 
all spectra with [Fe/H]$<-1$ and [Fe/H]$>-1$, respectively. The co-added
spectra are shown in Figure~\ref{fig:sp_coadd} and the corresponding
datapoints are shown with filled circles in Fig.~\ref{fig:popmod}. 
In the H$\beta$ plots both the metal-poor and metal-rich co-added datapoints 
fall near the $\sim11$ Gyr Maraston isochrones, with little evidence of any 
significant age differences between the two within the formal $\sim2$ Gyr 
error bars on the co-added data. The Schiavon models tend to
predict lower ages for a given H$\beta$ index than the Maraston models, 
especially at younger ages and lower metallicities. At the location of the 
metal-rich Sombrero GCs in the \fe -- H$\beta$ plane the difference is 
relatively modest and amounts to an age difference of only 1--2 Gyrs, but at 
lower metallicities the difference between the two sets of models becomes
more pronounced and amounts to about 4 Gyrs at the lowest metallicity
tabulated by Schiavon. Thus, the trend in these models is to decrease the 
ages of the metal-poor clusters more than those of the metal-rich ones, 
relative to the Maraston models.

  The H$\gamma_A$ and H$\delta_A$ measurements basically confirm the
ages estimated from H$\beta$, but note that both H$\gamma_A$ and H$\delta_A$
reach a minimum around 11 Gyr at low metallicities and then increase strongly
at higher ages.  At $\feh\ = -1.35$,
the Maraston models predict about the same H$\gamma_A$ and H$\delta_A$
strengths at 15 and 5 Gyrs.  The Schiavon models do not include
tabulations of H$\gamma_A$ and H$\delta_A$, but in 
Figure~\ref{fig:hgv_fe} we show their predictions for another H$\gamma$
index defined by \citet{vaz01}. This H$\gamma$ index (in the following
denoted H$\gamma_{\rm vaz}$) differs from H$\gamma_A$
in that the continuum passbands are located closer to the feature passband,
thereby avoiding the strong metallicity dependence of H$\gamma_A$ which 
results from its blue continuum passband coinciding with the G--band.
In fact, Figure~\ref{fig:hgv_fe} shows that H$\gamma_{\rm vaz}$ is
practically independent of metallicity, at least within the range spanned
by the models. Taken at face value, Figure~\ref{fig:hgv_fe} would suggest
that the metal-rich clusters are slightly \emph{older} than the metal-poor
ones, but both sub-populations are compatible with the same (high)
age within the errors. Thus, although it should be stressed that even relative 
ages may still be somewhat model-dependent, it seems reasonable to conclude
that both GC sub-populations in the Sombrero are old, and probably coeval
to within the measurement uncertainties, i.e.\ with age differences of
2--3 Gyrs at most.

\subsection{Abundance ratios}
\label{sec:alpha}

  Some attempts to model the effect of $\alpha$-element enhancement on Lick 
indices have been made by \citet{mbs00}, who modeled the effect of 
$\alpha$-element enhancement on the Lick \mgtwo\ index and a broad index 
(TiO$_{12.5}$) covering the TiO band centered at 7257 \AA . The TiO$_{12.5}$ 
index is included in our red LRIS spectra.  Because of the problems with 
the calibration of \mgtwo , we converted the Milone et al.\ model predictions 
for \mgtwo\ to \mgb , using the Maraston models. We found that the linear 
relation $\mgb = 14.03 \, \mgtwo\ + \, 0.361$ approximated the Maraston 
models to within 0.1 \AA\ for the age range 10--15 Gyr and $\mgtwo < 0.3$
(Fig.~\ref{fig:mgb_mg2}).

  In addition to the strong terrestrial O$_2$ absorption bands, which 
are located just outside the TiO$_{12.5}$ continuum passbands, the wavelength
region around 7200--7500\AA\ is affected by terrestrial H$_2$O absorption.
We corrected our red spectra for this by normalizing the flux standard
spectra with a smooth continuum and then dividing the GC spectra by the 
normalized flux standard spectra. This correction affected the TiO$_{12.5}$ 
index downwards by about 7\AA.  Fig.~\ref{fig:fe_alpha} shows our measurements 
of TiO$_{12.5}$ and $\mgb$ on the co-added spectra, compared to the 
alpha-enhanced models 
by \citet{mbs00}. In the $\fe - \mgb$ plot we also show the Maraston models 
for an age of 12 Gyr. Note that the most metal-poor Maraston models are 
actually rather similar to the $\alphafe=+0.3$ models by \citet{mbs00},
but approach the $\alphafe=0$ models at higher metallicities. This is
probably because many of the metal-poor stars used to set up the
Worthey fitting functions are alpha-enhanced \citep{mt00}.  For the
metal-rich Sombrero GCs, both \mgb\ and TiO$_{12.5}$ indicate clearly
supersolar \alphafe\ ratios around $\alphafe\sim+0.4$. For the metal-poor
clusters, the TiO$_{12.5}$ index again indicates supersolar \alphafe\ 
ratios similar to those of the metal-rich clusters, although the \mgb\ 
plot actually suggests \alphafe\ closer to 0 for the metal-poor clusters.
However, at low metallicities the separation between the different 
\alphafe\ models is smaller and it seems most likely that both the
metal-poor and metal-rich GCs in the Sombrero have super-solar $\alphafe$
ratios similar to those observed in Milky Way GCs, the bulge of the
Sombrero and in early-type galaxies.

\section{Mass of the Sombrero}
\label{sec:mass}

  Globular clusters can be used as tracers of the mass of their host galaxy,
as demonstrated e.g.\ by \citet{hb87} (for M87) and \citet{kis98} (NGC 1399).
We combined our data in Table~\ref{tab:clusters} with data for 34
Sombrero GCs from \citet{bri97}. Our sample does not overlap with that
of \citet{bri97}, so by combining theirs and our data we get a total of 48 GCs.

  Fig.~\ref{fig:rvplot} shows the radial velocities as a function of the
offset in R.A. from the center of the Sombrero (the major axis is aligned
roughly east-west). \citet{bri97} found hints of rotation in the cluster
system, but our data clearly do not add much to the significance of their
result. For completeness we also plot the radial velocities for the metal-rich 
and metal-poor clusters from our sample separately, but the data are clearly 
insufficient to provide any constraints on rotation in either sample.

  Following \citet{hb87}, we can use the GC radial velocities to obtain a 
mass estimate for the Sombrero in two ways. The virial mass is estimated as
\begin{equation}
  M_{\rm VT} \; = \; \frac{3\pi N}{2G} 
            \frac{\Sigma_i^N \, V_i^2}{\Sigma_{i<j}1/r_{ij}}
\end{equation}
and the projected mass is
\begin{equation}
  M_P \; = \; \frac{f_p}{GN} \left( \Sigma_i^N V_i^2 r_i \right)
\end{equation}
  In these equations, $r_{ij}$ is the distance between the $i$th and
$j$th clusters, $r_i$ is the distance between the $i$th cluster and the
galaxy center and $V_i$ is the velocity difference between the $i$th cluster
and the mean system velocity. $N$ is the number of clusters. The 
value of $f_p$ depends on the cluster orbits; following \citet{bri97} we 
adopt $f_p = 16/\pi$ for isotropic orbits.

  For the total sample of 48 clusters we get 
$M_{\rm VT} \, = \, (8.1\pm1.8)\times10^{11} \, \msun$ and
$M_P \, = \, (5.3\pm1.0)\times10^{11} \, \msun$ within 17 kpc, adopting 
a distance of 8.7 Mpc \citep{lfb01}. Using only our 14 clusters, we get
$M_{\rm VT} \, = \, (3.2\pm1.5)\times10^{11} \, \msun$ and
$M_P \, = \, (4.2\pm2.1)\times10^{11} \, \msun$ (within 13 kpc).  For 
comparison, \citet{bri97} got $M_P \, = \, 5.2^{6.7}_{3.9}\times10^{11}\msun$.
The 12 clusters in the central HST pointing are useful as probes of the 
mass located within the central regions of the galaxy, and yield
$M_{\rm VT} \, = \, (2.1\pm1.1)\times10^{11} \, \msun$ and 
$M_P \, = \, (2.0\pm1.3)\times10^{11} \, \msun$ within the central 4.5 kpc.
The errors 
were estimated by a standard ``jackknife'' procedure.  

\section{Summary and conclusions}

  We have analyzed Keck / LRIS-B spectra for 17 globular cluster
(GC) candidates in the Sombrero galaxy, of which 14 are confirmed as true GCs 
based on their radial 
velocities. One additional object has too low S/N for reliable radial 
velocity measurements, but based on the HST images presented by \citet{lfb01} 
it is probably also a GC.  We find generally good agreement between 
spectroscopic metallicities and previous photometric metallicites based on 
\vi\ colors. Comparison of Lick/IDS indices with population synthesis models
indicates that both the metal-rich and metal-poor clusters are 10--15 Gyr
old, with preferred ages around 11-12 Gyr. We see no 
detectable age differences between the two GC populations, although 
differences of a few Gyrs cannot be ruled out.  Comparison of \fe, \mgb\ 
and TiO indicates that the Sombrero GCs probably have super-solar 
\alphafe\ ratios $\sim+0.4$, similar to or perhaps slightly larger than
those observed in Milky Way globular clusters.

  Overall, the Sombrero GC system shares many similarities with that of
the Milky Way, as well as many early-type galaxies. With few exceptions,
both GC populations generally appear to be old, and another common feature 
seems to be an enhancement in $\alpha$-element abundances, probably 
indicating that the gas out of which the clusters formed was enriched 
predominantly by type-II supernovae. This suggests that the majority of 
GCs formed early on, before type Ia SNe made a significant contribution to 
chemical enrichment, and presumably within a few Gyrs after star formation 
commenced.  The metal-rich GCs in the Sombrero, like those in most 
galaxies, probably formed early on when galaxies were still assembling with
high star formation rates and gas densities providing favorable conditions
for the formation of large numbers of massive star clusters. The metal-poor
clusters might have formed even earlier at high redshifts, triggered by 
ionization fronts \citep{cen01}.

  Finally, we combine our sample with data for 34 GCs from \citet{bri97} and 
obtain virial and projected masses for the Sombrero of 
$M_{\rm VT} \, = \, (8.1\pm1.8)\times10^{11} \, \msun$ and
$M_P \, = \, (5.3\pm1.0)\times10^{11} \, \msun$, respectively. This is
in good agreement with previous estimates of the mass of the Sombrero.
For 12 clusters located within the central 4.5 kpc (projected) we
estimate $M_{\rm VT} \, = \, (2.1\pm1.1)\times10^{11} \, \msun$ and
$M_P \, = \, (2.0\pm1.3)\times10^{11} \, \msun$.

\acknowledgments

  This work was supported by National Science Foundation grant number 
AST9900732. We are grateful to Dr.\ Andre Milone for providing his 
$\alpha$-enhanced models in electronic form and to Dr.\ Ricardo Schiavon for 
allowing us to use his population models prior to publication. Helpful
comments by an anonymous referee are appreciated.  MB thanks 
the Royal Society of London for their support.

\newpage
\onecolumn

\epsfxsize=16cm
\epsfbox{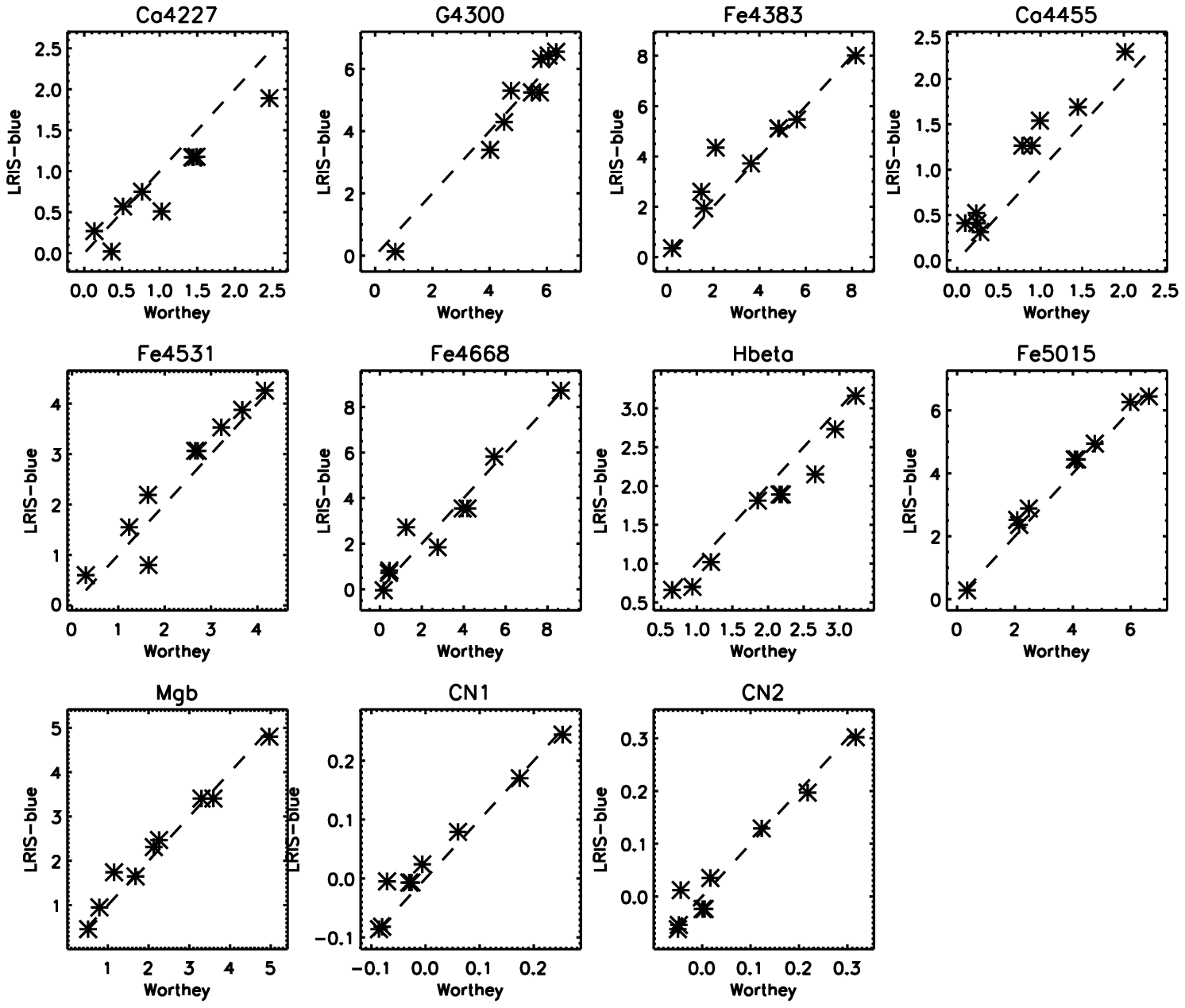}
\figcaption[Larsen.fig1.ps]{\label{fig:stdcmp}Lick/IDS indices measured on 
our LRIS-B spectra, compared to standard values from \citet{wor94}. The 
agreement is generally good, except for narrow indices like Ca4227 and Ca4455.}

\epsfxsize=16cm
\epsfbox{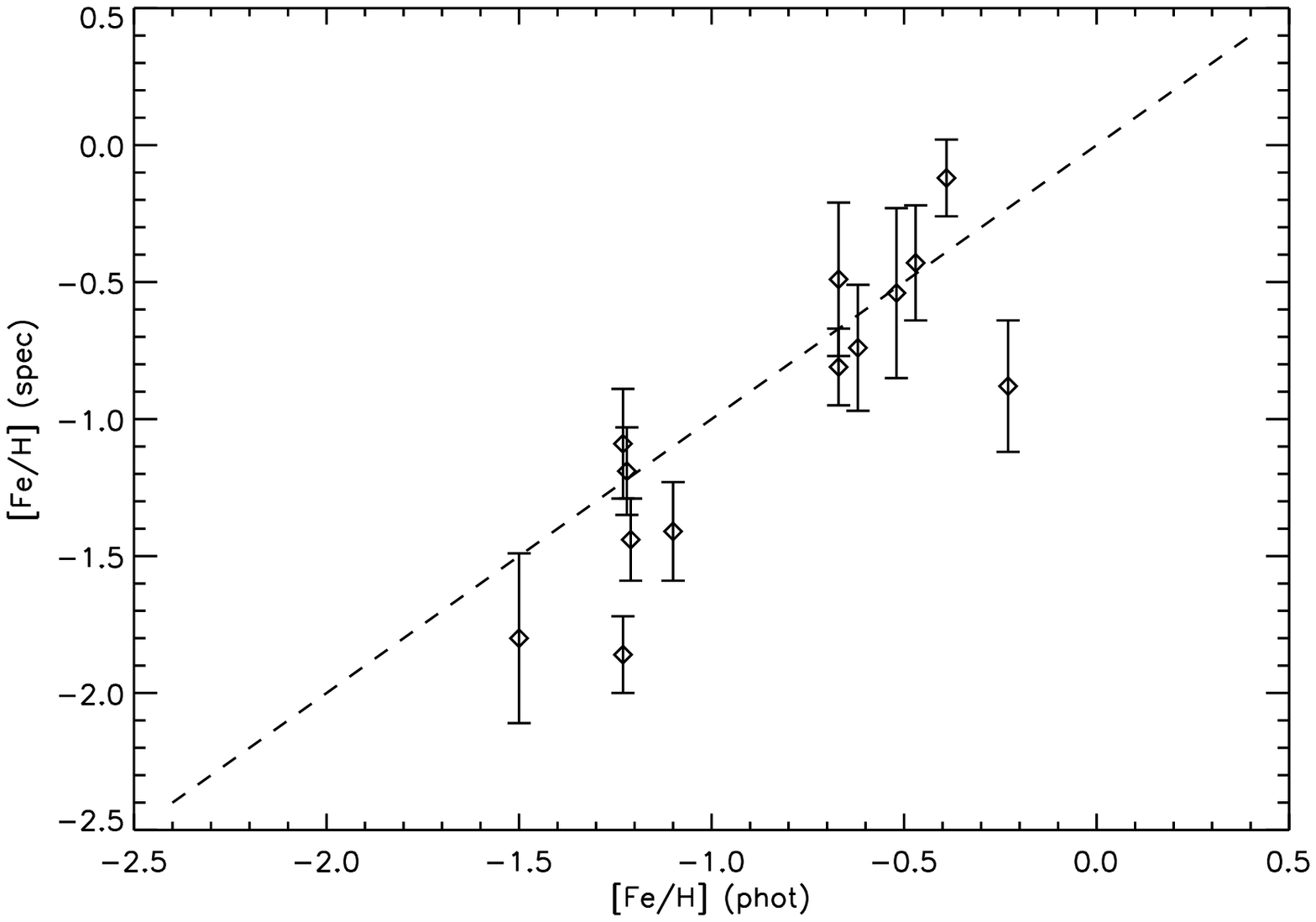}
\figcaption[Larsen.fig2.ps]{\label{fig:metcmp}Comparison of spectroscopic 
and photometric metallicity estimates for Sombrero GCs.}
\newpage

\epsfxsize=16cm
\epsfbox[90 370 540 700]{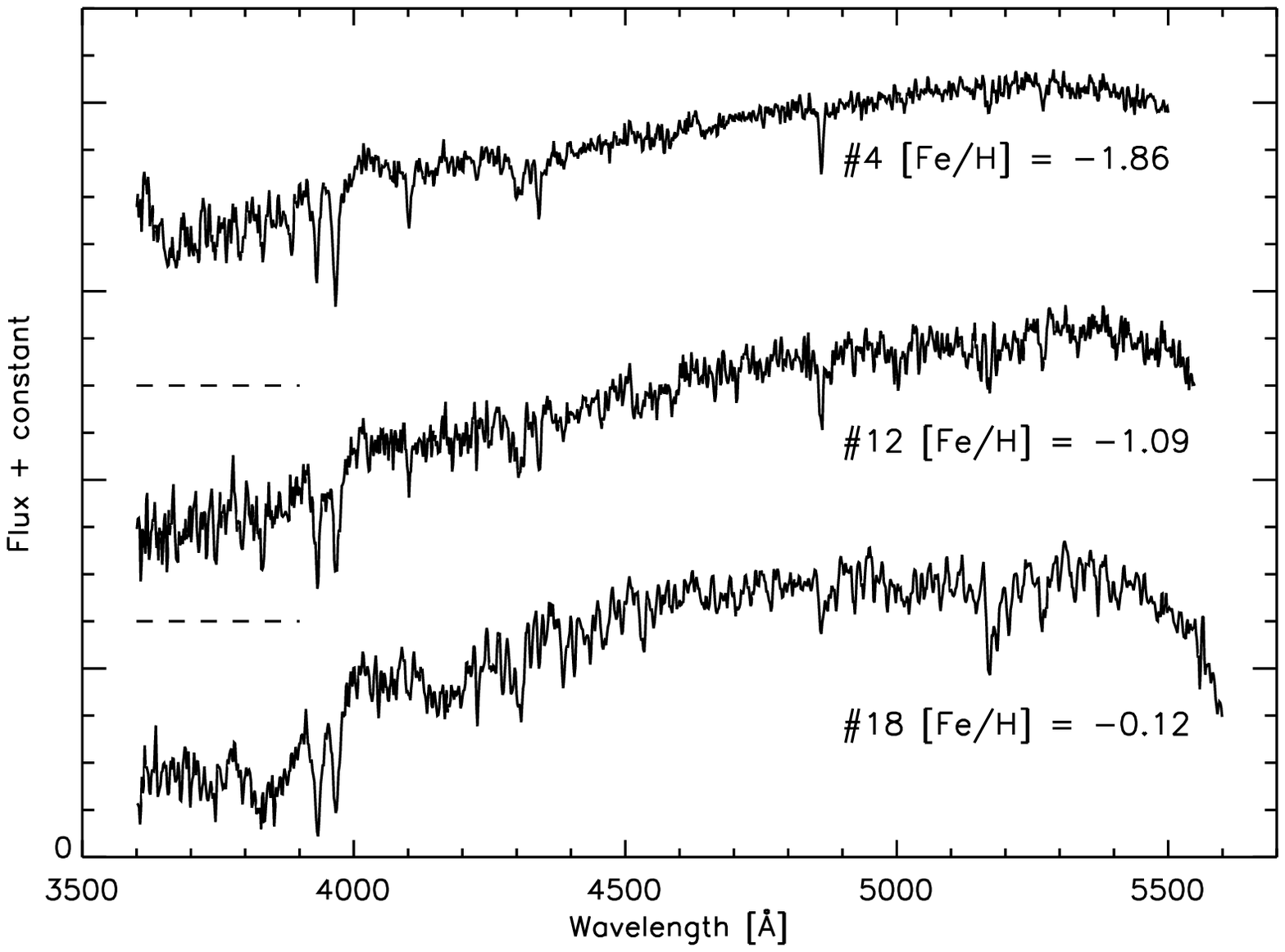}
\figcaption[Larsen.fig3.ps]{\label{fig:spectra}Spectra for three globular 
clusters with low (top), moderate (center) and high (bottom) metallicities. 
Note: no smoothing has been done. The two upper spectra have been shifted
for clarity, with the zero-levels indicated by the short dashed lines.
}
\newpage

\epsfxsize=16cm
\epsfbox{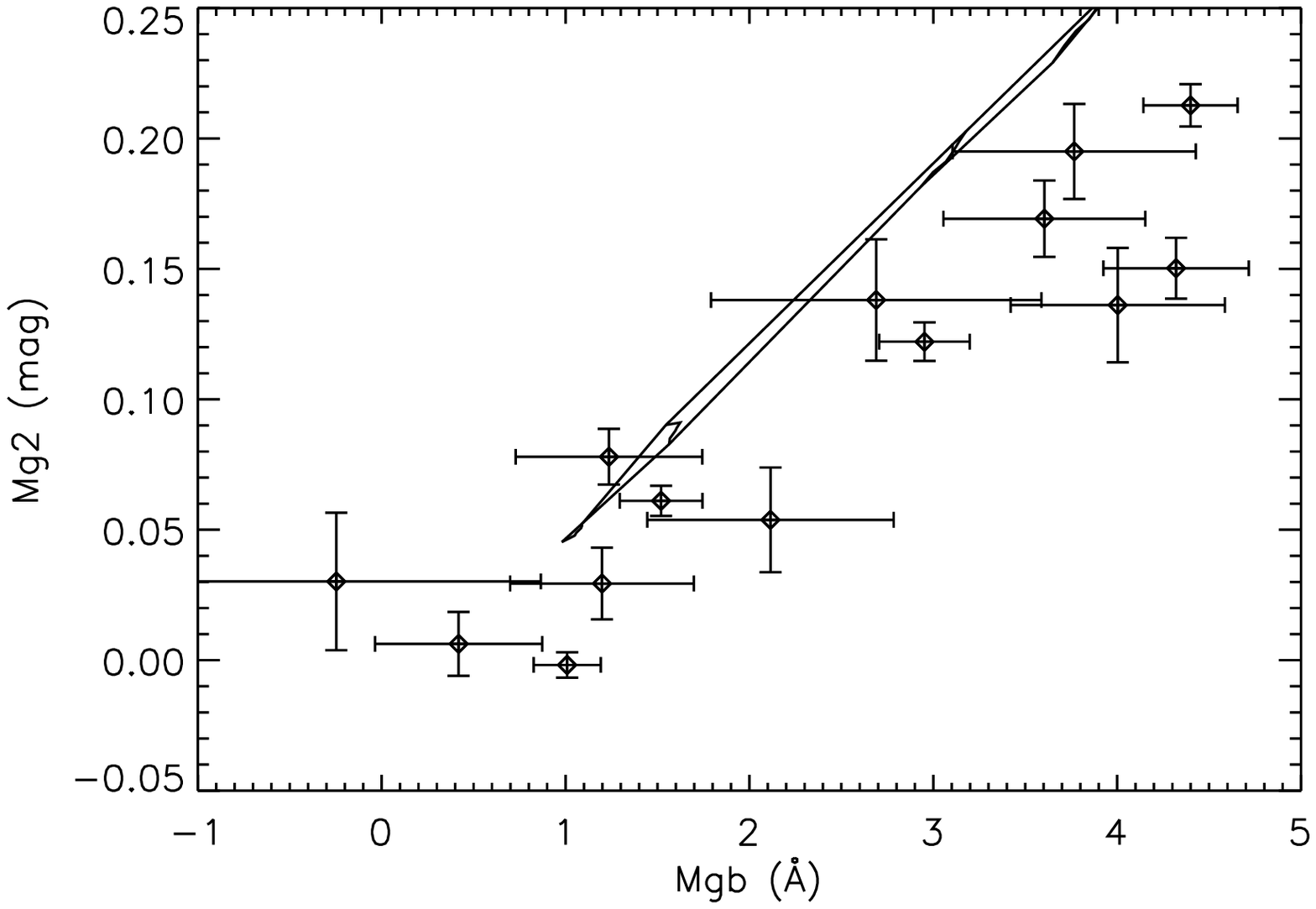}
\figcaption[Larsen.fig4.ps]{\label{fig:mgb_mg2}Comparison of 
\mgb\ and \mgtwo\ for Sombrero GCs.  Maraston models for 10 and 15 Gyr 
are overplotted on the figure.}
\newpage

\begin{flushleft}
\begin{minipage}{16cm}
\begin{minipage}{8cm}
\epsfxsize=8cm
\epsfbox{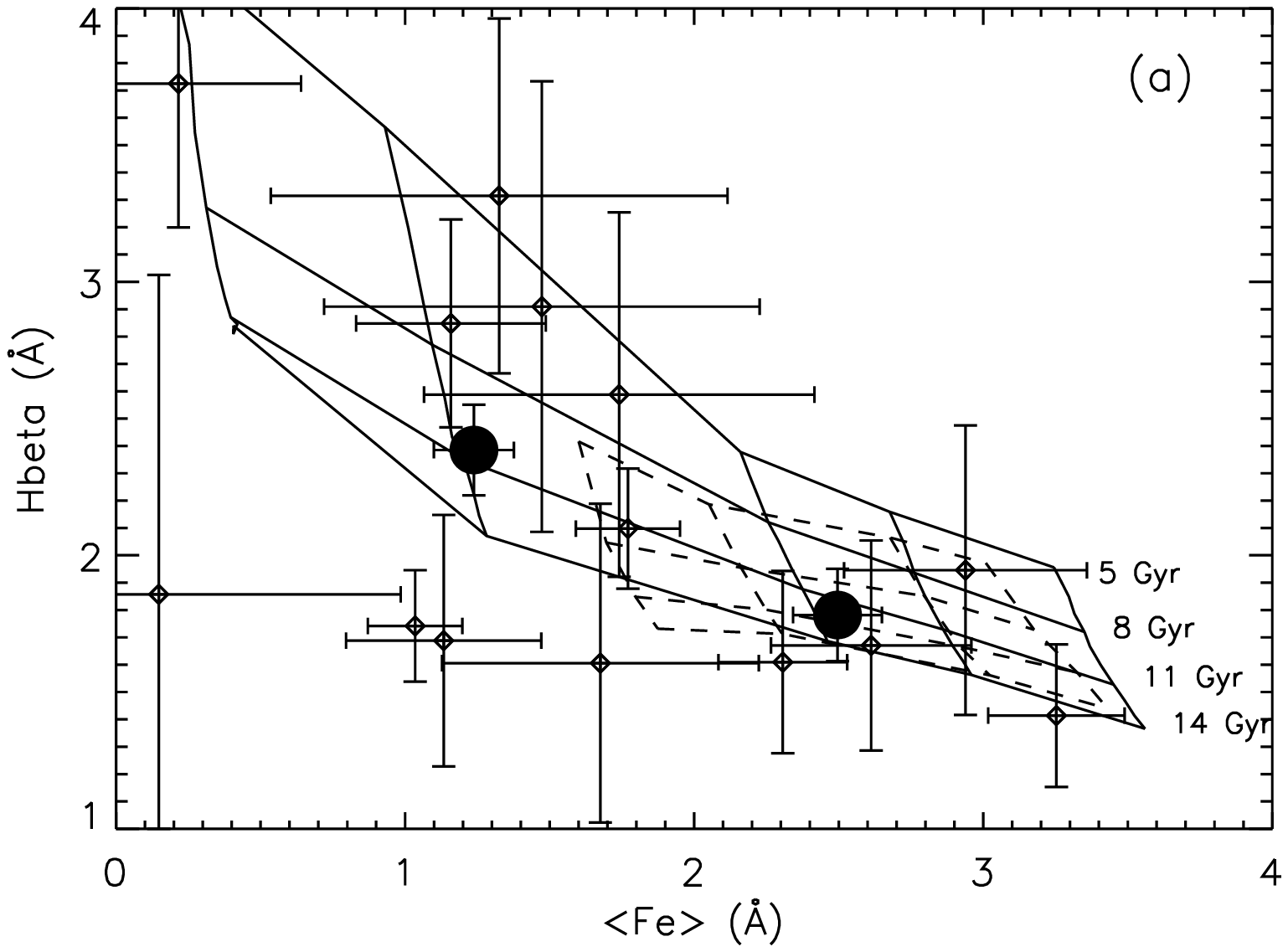}
\end{minipage}
\begin{minipage}{8cm}
\epsfxsize=8cm
\epsfbox{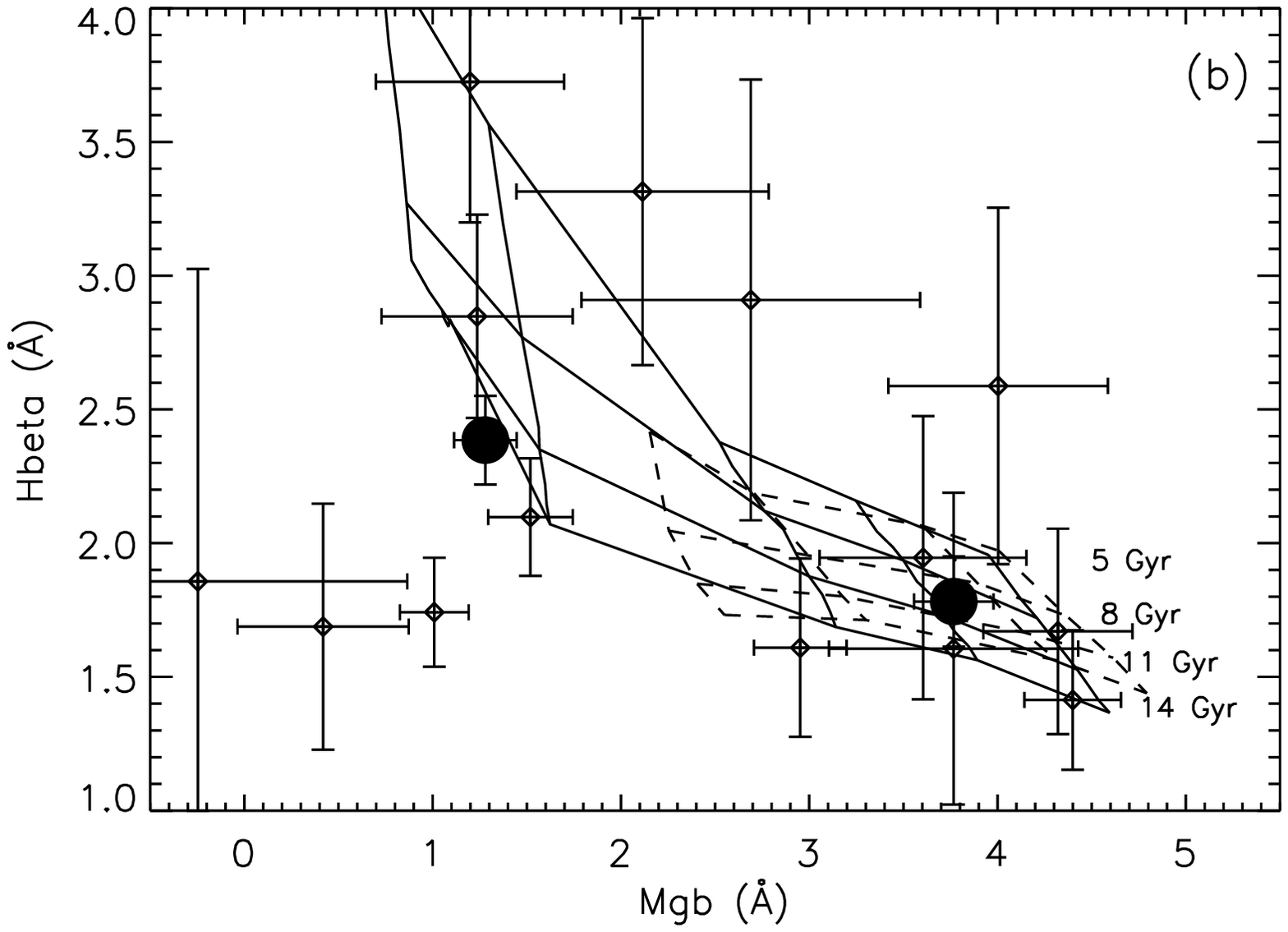}
\end{minipage}
\end{minipage}
\\
\begin{minipage}{16cm}
\begin{minipage}{8cm}
\epsfxsize=8cm
\epsfbox{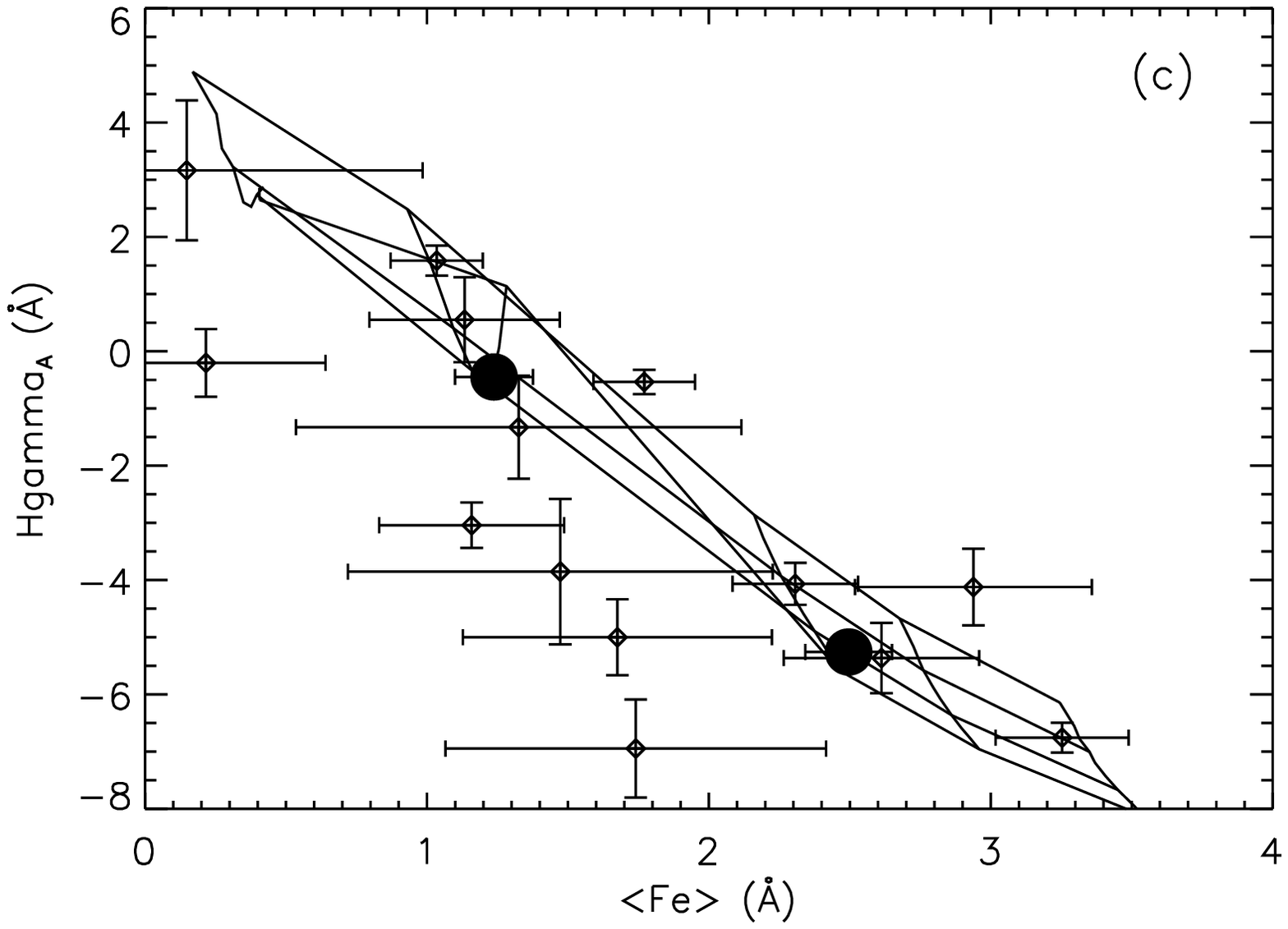}
\end{minipage}
\begin{minipage}{8cm}
\epsfxsize=8cm
\epsfbox{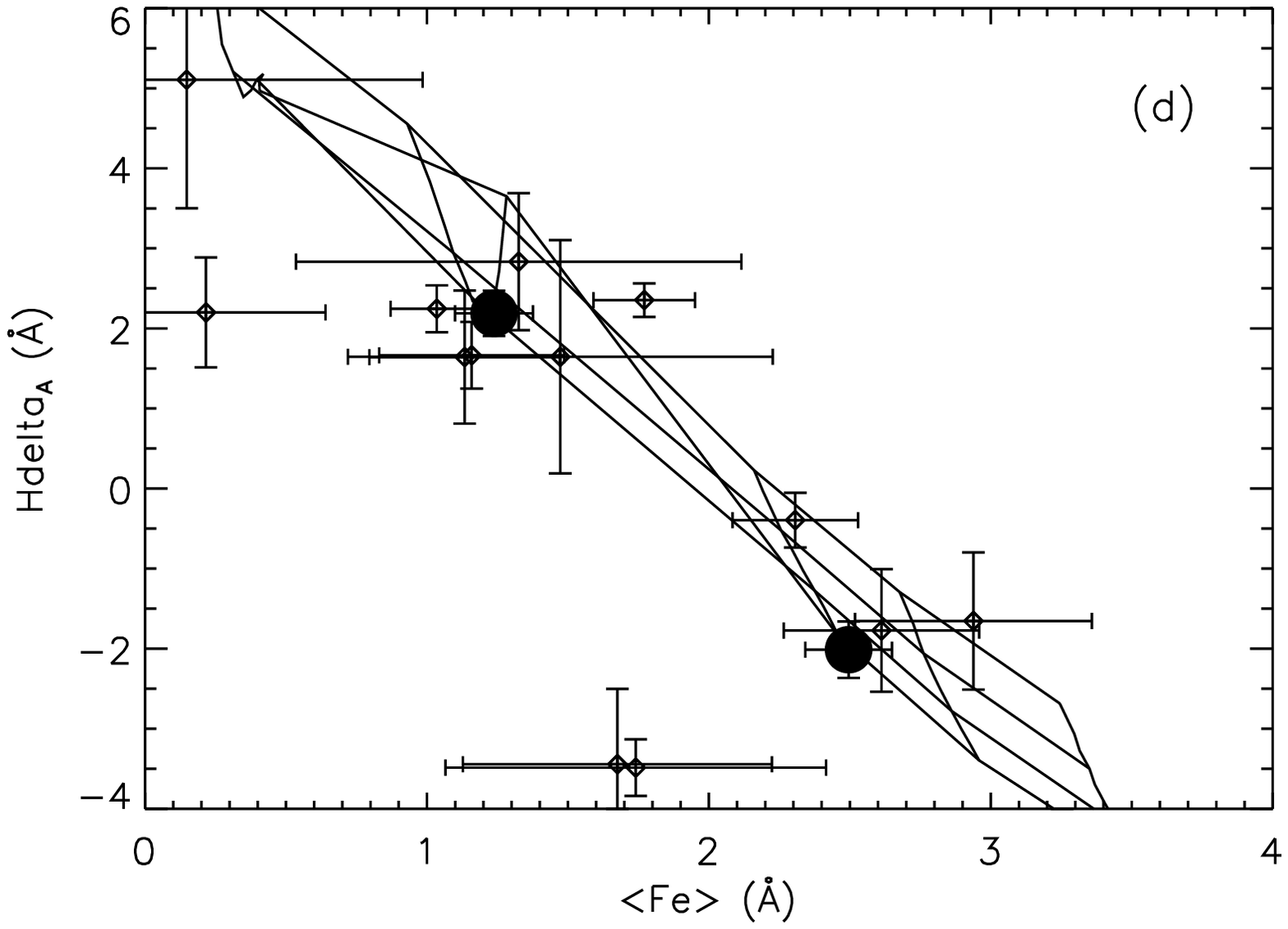}
\end{minipage}
\end{minipage}
\end{flushleft}
\figcaption[Larsen.fig5a.ps,Larsen.fig5b.ps,Larsen.fig5c.ps,Larsen.fig5d.ps]
{\label{fig:popmod}Comparison of our data with population
synthesis models by C.\ Maraston and R.\ Schiavon. 
Panels (a) and (b): \fe\ and \mgb\ vs.\ H$\beta$. 
Panels (c) and (d): \fe\ vs.\ H$\gamma_A$ and H$\delta_A$. Maraston models
(solid line grids) are shown for ages of 5, 8, 11 and 14 Gyr and
\feh\ = $-2.25$, $-1.35$, $-0.33$, 0.0 and $+0.35$. Schiavon models
(dashed line grids) are for ages of 5, 7.9, 11.2 and 14.1 Gyr and 
\feh\ = $-0.7$, $-0.4$, 0.0 and $+0.2$. The filled circles represent
the co-added metal-poor and metal-rich spectra.
}
\newpage

\epsfxsize=12.5cm
\noindent \epsfbox{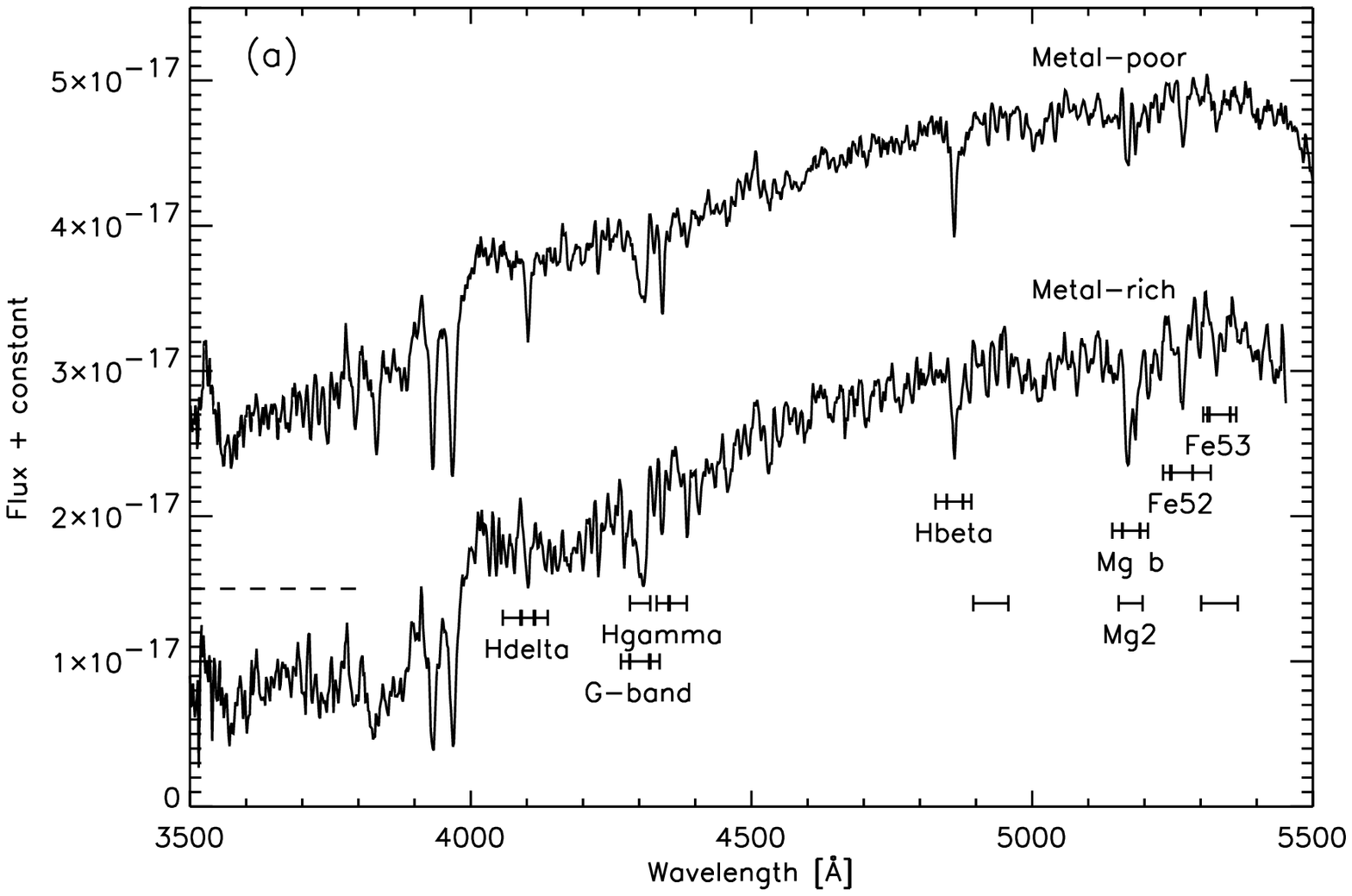} \\
\epsfxsize=12.5cm
\noindent \epsfbox{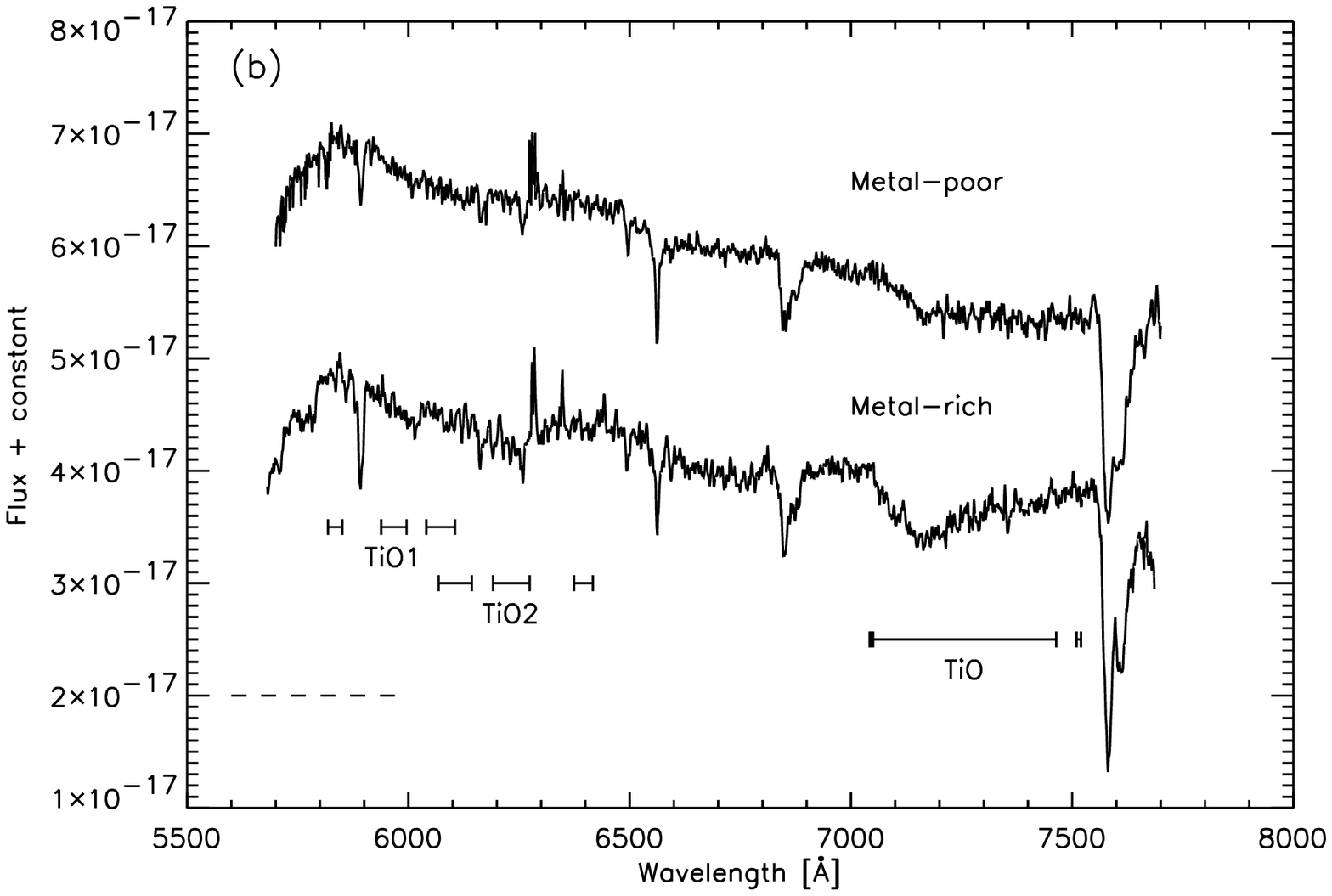}
\figcaption[Larsen.fig6a.ps,Larsen.fig6b.ps]
  {\label{fig:sp_coadd}Co-added spectra of all metal-rich
($\feh > -1.0$) and metal-poor ($\feh < -1.0$) globular clusters. 
Panels (a) and (b) show the spectra from the blue and red side arms on
LRIS, respectively.  Lick indices used in this paper are indicated.
The metal-poor spectra have been shifted upwards, as indicated by 
the horizontal dashed lines.
}
\newpage

\epsfxsize=16cm
\epsfbox{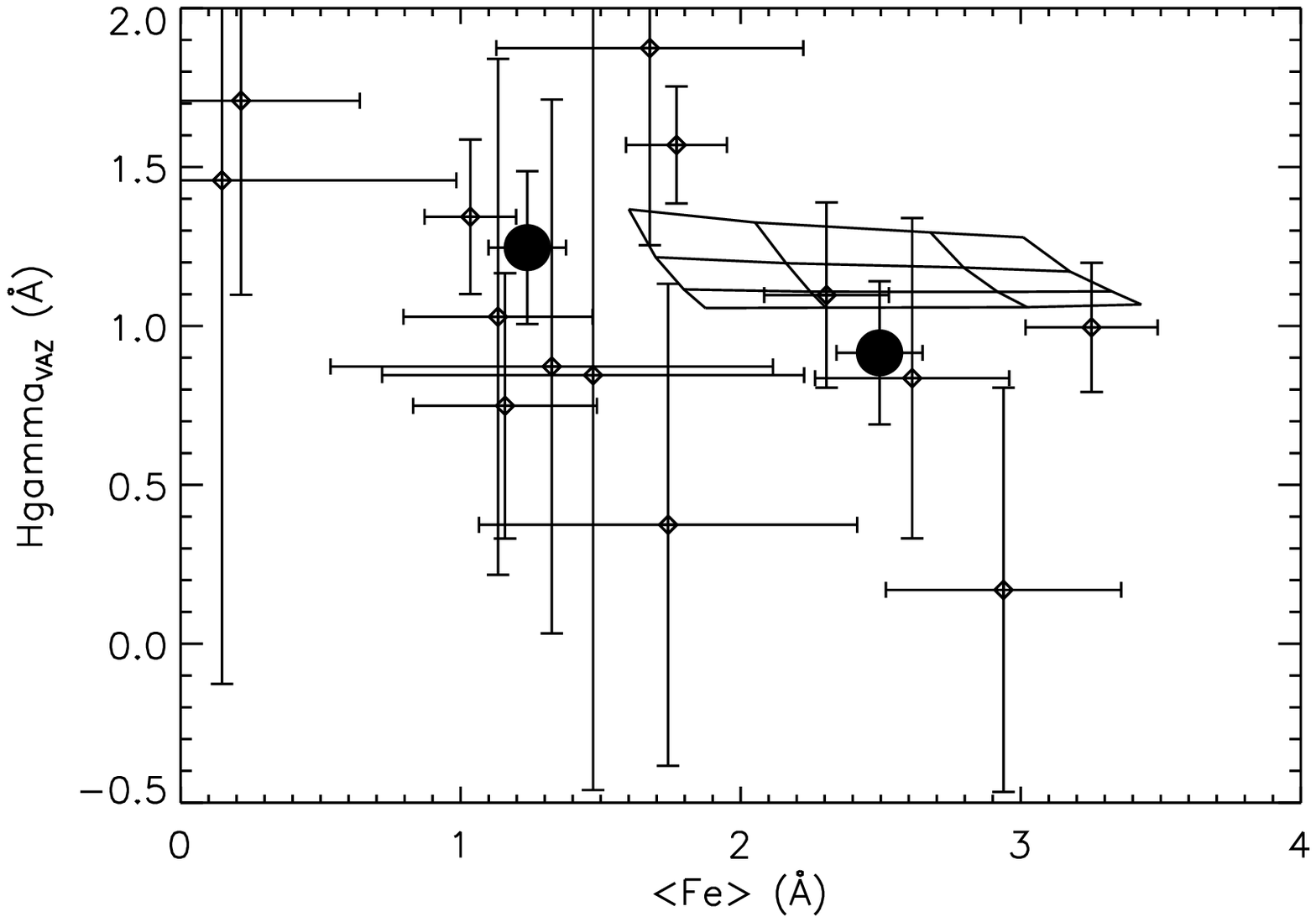}
\figcaption[Larsen.fig7.ps]{\label{fig:hgv_fe}Measurements of
H$\gamma_{\rm vaz}$ according 
to the \citet{vaz01} definition and \fe, compared to Schiavon (2002) models.
}

\begin{flushleft}
\epsfxsize=12cm
\epsfbox{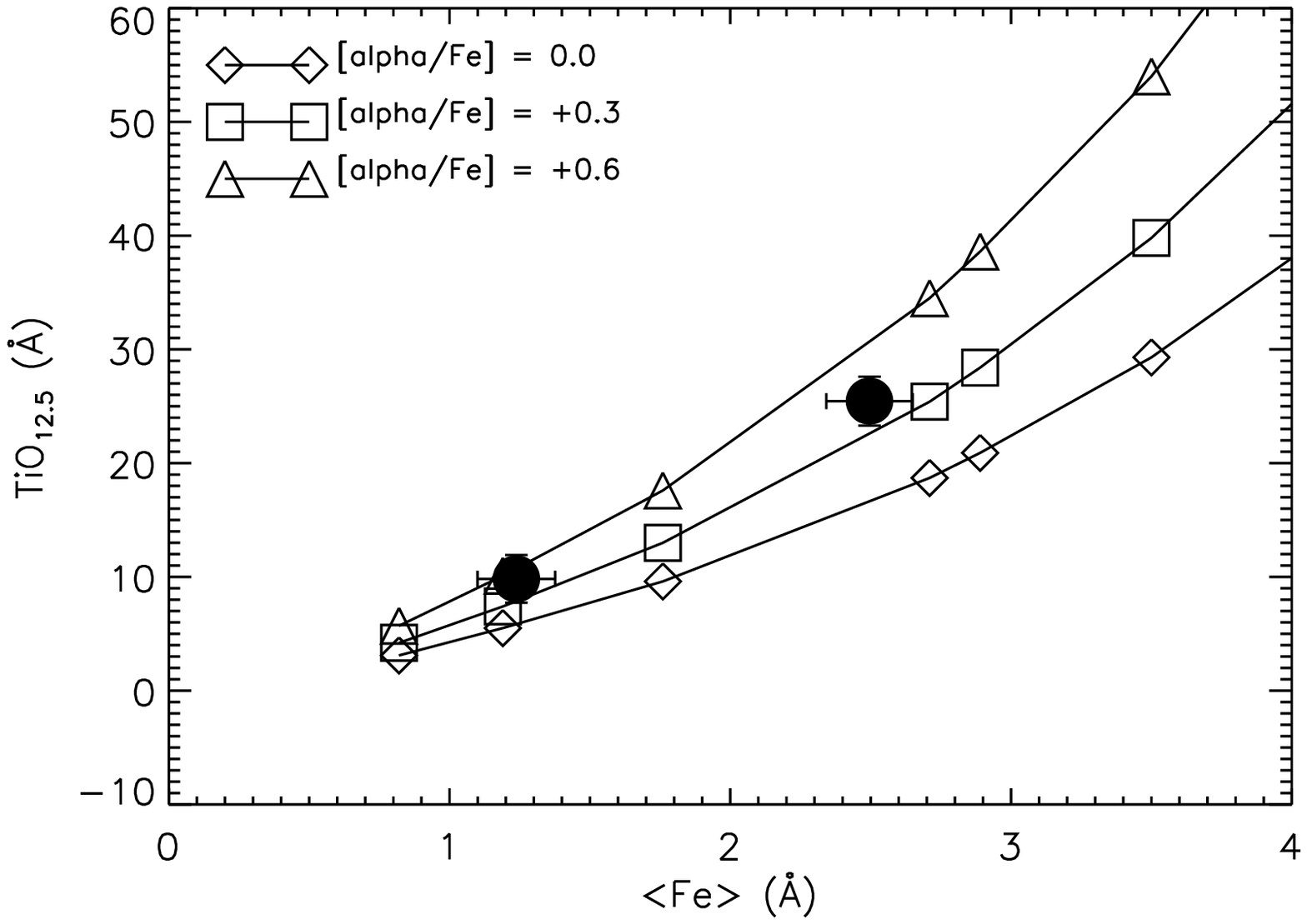} 

\epsfxsize=12cm
\epsfbox{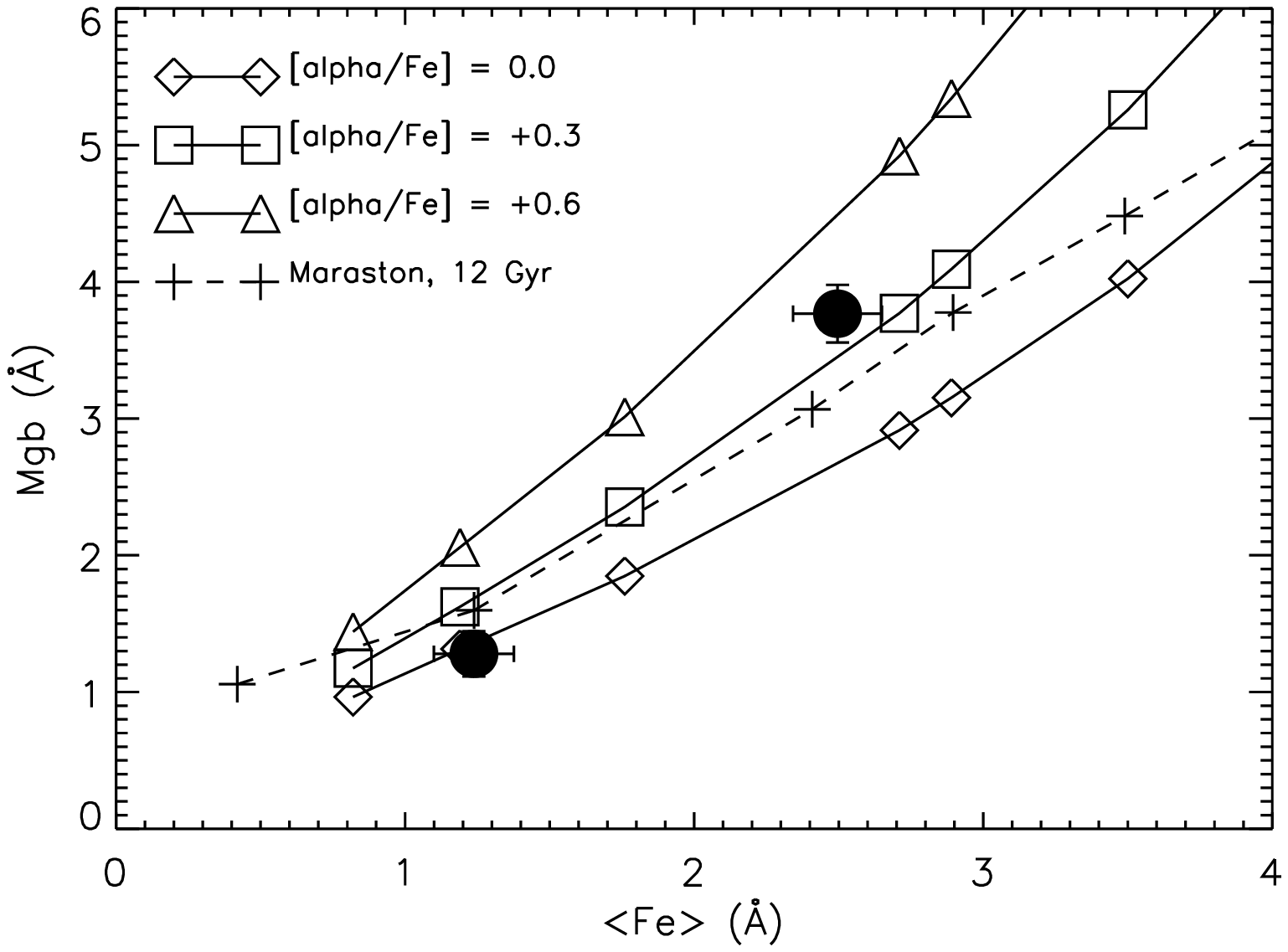}
\end{flushleft}
\figcaption[Larsen.fig8a.ps,Larsen.fig8b.ps]{\label{fig:fe_alpha}TiO$_{12.5}$ 
and \mgb\ vs.\ \fe\ for
co-added metal-rich and metal-poor spectra, compared to $\alpha$-enhanced
models by \citet{mbs00}. The dashed line in the \mgb\ vs.\ \fe\ plot
represents 12 Gyr Maraston models.}
% \newpage

\epsfxsize=15cm
\epsfbox{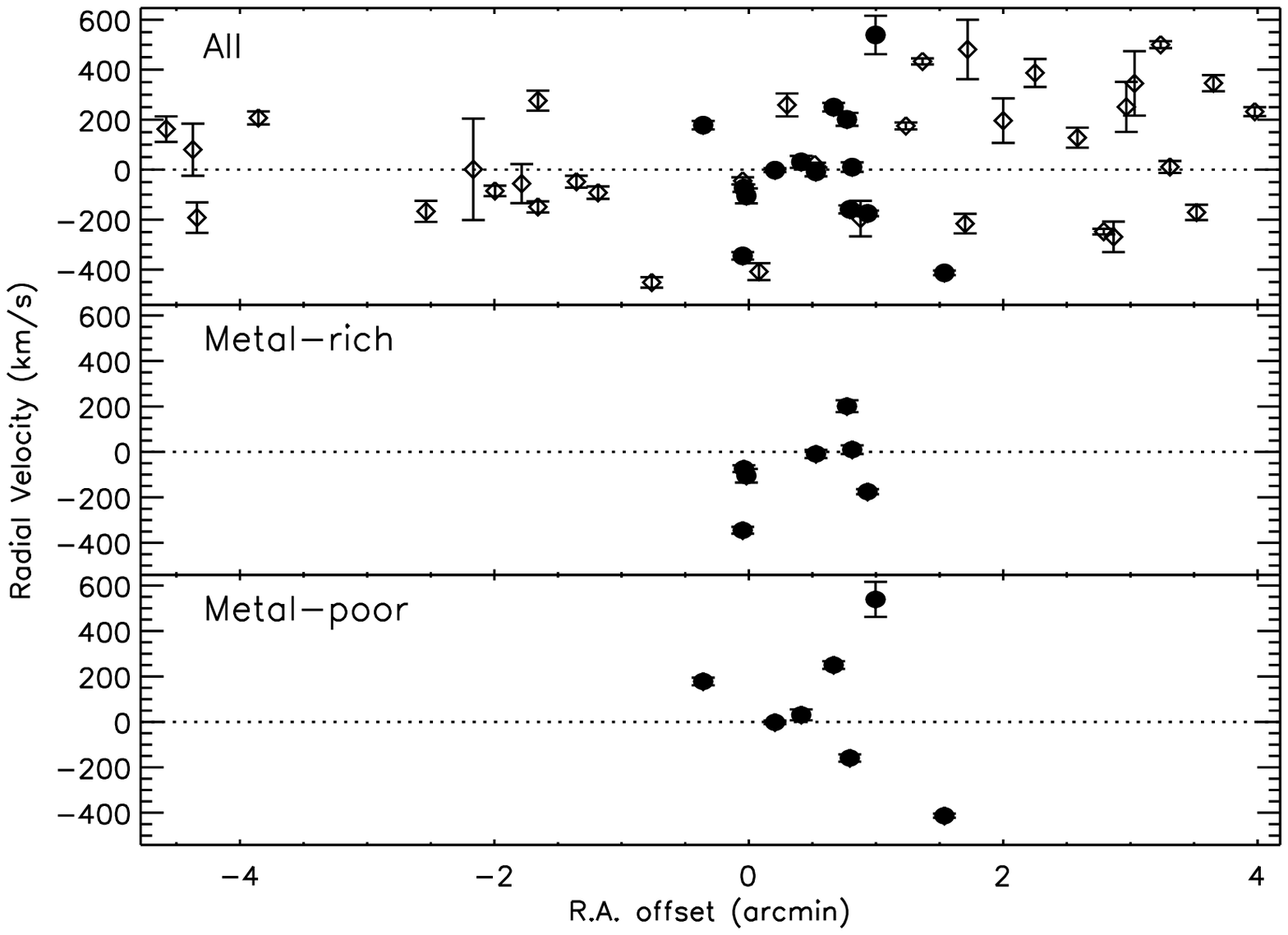}
\figcaption[Larsen.fig9.ps]{\label{fig:rvplot}Radial velocity as a function 
of distance from the center along the major axis for our full GC sample
combined with data from \citet{bri97} (top), 
our metal-rich clusters only (center) and our metal-poor clusters only 
(bottom). Data
from \citet{bri97} are shown with open circles while our data are shown
with filled circles. No significant evidence for rotation is seen in 
either sample.}
% \newpage

\begin{deluxetable}{lcccrrl}
\tabletypesize{\small}
\tablecaption{ \label{tab:clusters}Globular cluster candidates in NGC~4594}
\tablecomments{RV = heliocentric radial velocity,
  $V$ = visual magnitude.  The S/N column gives the average signal-to-noise 
  per pixel in the region 4500\AA -- 5000\AA\ (blue) and 6500\AA -- 7000\AA 
  (red).  $^a$: This object was saturated on the WFPC2 data used in Paper I.
  $^b$: Radial vel.\ from blue spectrum. The IDs refer to \citet{lfb01},
  where coordinates are listed. The quoted errors on the radial velocities
  are random errors determined from the uncertainty on the cross-correlation
  peaks as reported by FXCOR.  Additional systematic errors are $\sim25$ km/s 
  (see text).
}
\tablehead{
   ID      & $V$   & \viz  & RV     & \multicolumn{2}{c}{S/N} & GC?\\
           &       &       & km/s   & Blue & Red   & }
\startdata
1  / H2-06 & 20.28 &  0.92 & $ 304\pm 11^b$ &  46 & -  & N \\
2  / H2-10 & 20.51 &  1.01 & $ 108\pm  8$ &  45 & 55   & N \\
% 3  /   -   -     &   -   &   -   & $ 232\pm 11$ & 129 & 110 \\
4  / H2-09 & 20.45 &  0.97 & $ 611\pm  9$ &  45 & 44   & Y \\
5  / H2-22 & 19.1$^a$ &  -  & $1274\pm 17$ &  19 & 24  & Y \\
6  / H2-27 & 22.04 &  1.14 & $ 227\pm 39$ &   9 & 9    & Y? \\
7  / C-076 & 21.36 &  0.90 & $1563\pm 77^b$ & 8 & 16   & Y \\
8  / C-068 & 20.37 &  1.11 & $ 849\pm 11$ &  32 & 41   & Y \\
9  / C-051 & 21.17 &  1.18 & $1015\pm 18$ &  17 & 25   & Y \\
10 / C-064 & 21.10 &  0.94 & $ 865\pm 16$ &  17 & 23   & Y \\
11 / C-059 & 21.14 &  1.17 & $1034\pm 19$ &  13 & 22   & Y \\
12 / C-032 & 20.38 &  0.97 & $1055\pm 24$ &  25 & 36   & Y \\
13 / C-042 & 21.19 &  1.26 & $1225\pm 26$ &  11 & 15   & Y \\
% 14 / Nucleus &   -       &      -      &   -   &   -   & $1637\pm 50$ &  28 & 56 \\
15 / C-116 & 19.98 &  0.96 & $1022\pm  8$ &  46 & 46   & Y \\
16 / C-136 & 21.53 &  1.06 & $1202\pm 17$ &  14 & 14   & Y \\
17 / C-132 & 20.91 &  1.15 & $ 679\pm 15$ &  16 & 23   & Y \\
18 / C-134 & 20.29 &  1.24 & $ 950\pm 15$ &  41 & 34   & Y \\
19 / C-137 & 21.10 &  1.19 & $ 919\pm 30$ &  24 & 33   & Y \\
\enddata
\end{deluxetable}

\begin{deluxetable}{lrrrrrrrrrr}
\rotate
\tablewidth{0pt}
\tabletypesize{\tiny}
\tablecaption{Metallicity estimates based on the calibration in
  \citet{bh90}. The two bottom lines give data for the co-added metal-poor
  ($\feh<-1$) and metal-rich ($\feh\ge-1$) spectra.
\label{tab:mettab}
}
\tablehead{
Object   & \multicolumn{8}{c}{Individual [Fe/H] estimators} &
          [Fe/H] (sp)    &  [Fe/H] (phot) \\
           & $\Delta$ & \mgtwo\ &  MgH  & Gband &  CNB &  Fe5270 &  CNR & H$+$K
   }
\startdata
4 & $-1.59\pm 0.37$ & $-2.29\pm 0.34$ & $-2.35\pm 0.48$ & $-1.72\pm 0.33$ & $-1.96\pm 0.34$ & $-1.57\pm 0.62$ & $-1.45\pm 0.45$ & $-1.72\pm 0.43$ & $-1.86\pm 0.14$ &  -1.23 \\
5 & $-1.28\pm 0.37$ & $-2.23\pm 0.36$ & $-2.06\pm 0.52$ & $-1.07\pm 0.41$ & $-1.34\pm 0.41$ & $-1.63\pm 0.65$ & $-1.30\pm 0.49$ & $-0.19\pm 0.51$ & $-1.41\pm 0.27$ &  -0.42 \\
7 & $-0.87\pm 0.39$ & $-2.10\pm 0.43$ & $-1.30\pm 0.67$ & $-1.85\pm 0.64$ & $-0.78\pm 0.97$ & $-1.98\pm 0.82$ & $-2.40\pm 0.54$ & $-2.92\pm 1.28$ & $-1.80\pm 0.31$ &  -1.50 \\
8 & $-0.47\pm 0.37$ & $-1.10\pm 0.34$ & $-1.17\pm 0.49$ & $-0.50\pm 0.35$ & $-0.97\pm 0.36$ & $-0.90\pm 0.63$ & $-0.96\pm 0.46$ & $-0.26\pm 0.45$ & $-0.81\pm 0.14$ &  -0.67 \\
9 & $ 0.04\pm 0.37$ & $-0.68\pm 0.37$ & $-0.73\pm 0.53$ & $-1.80\pm 0.43$ & $ 0.01\pm 0.48$ & $-0.60\pm 0.70$ & $-0.63\pm 0.48$ & $ 0.55\pm 0.53$ & $-0.54\pm 0.31$ &  -0.52 \\
10 & $-1.06\pm 0.37$ & $-2.11\pm 0.36$ & $-1.22\pm 0.54$ & $-1.33\pm 0.40$ & $-1.17\pm 0.35$ & $-1.60\pm 0.68$ & $-1.42\pm 0.47$ & $-1.44\pm 0.48$ & $-1.44\pm 0.15$ &  -1.21 \\
11 & $ 0.24\pm 0.37$ & $-0.84\pm 0.40$ & $-1.84\pm 0.59$ & $-0.86\pm 0.48$ & $-0.63\pm 0.59$ & $-0.58\pm 0.76$ & $-0.56\pm 0.47$ & $-0.56\pm 0.61$ & $-0.74\pm 0.23$ &  -0.62 \\
12 & $-0.90\pm 0.37$ & $-1.57\pm 0.35$ & $-1.24\pm 0.51$ & $-0.45\pm 0.35$ & $-1.42\pm 0.35$ & $-1.58\pm 0.65$ & $-1.24\pm 0.46$ & $-0.38\pm 0.45$ & $-1.09\pm 0.20$ &  -1.23 \\
13 & $-0.59\pm 0.38$ & $-1.03\pm 0.41$ & $-1.91\pm 0.60$ & $ 0.16\pm 0.68$ & $-1.05\pm 0.44$ & $-1.04\pm 0.83$ & $-0.73\pm 0.53$ & $-0.59\pm 0.66$ & $-0.88\pm 0.24$ &  -0.23 \\
15 & $-0.88\pm 0.37$ & $-1.62\pm 0.34$ & $-1.83\pm 0.49$ & $-1.08\pm 0.32$ & $-1.02\pm 0.34$ & $-1.17\pm 0.62$ & $-1.26\pm 0.45$ & $-0.59\pm 0.43$ & $-1.19\pm 0.16$ &  -1.22 \\
16 & $-1.08\pm 0.37$ & $-1.62\pm 0.39$ & $-2.34\pm 0.58$ & $-1.09\pm 0.46$ & $-1.43\pm 0.39$ & $-1.56\pm 0.81$ & $-1.32\pm 0.48$ & $-0.84\pm 0.53$ & $-1.41\pm 0.18$ &  -1.10 \\
17 & $-0.30\pm 0.37$ & $-0.32\pm 0.38$ & $-1.79\pm 0.55$ & $-0.52\pm 0.48$ & $-0.75\pm 0.42$ & $-1.02\pm 0.71$ & $ 0.20\pm 0.49$ & $ 0.36\pm 0.53$ & $-0.49\pm 0.28$ &  -0.67 \\
18 & $-0.20\pm 0.37$ & $-0.08\pm 0.35$ & $-0.76\pm 0.49$ & $-0.24\pm 0.33$ & $-0.17\pm 0.34$ & $ 0.33\pm 0.63$ & $ 0.42\pm 0.46$ & $-0.15\pm 0.43$ & $-0.12\pm 0.14$ &  -0.39 \\
19 & $-0.07\pm 0.37$ & $-0.78\pm 0.36$ & $-1.00\pm 0.51$ & $-0.71\pm 0.46$ & $ 0.24\pm 0.37$ & $-0.90\pm 0.66$ & $-0.47\pm 0.47$ & $ 0.11\pm 0.47$ & $-0.43\pm 0.21$ &  -0.47 \\
metal-poor &  $-1.11\pm 0.37$ &  $-1.82\pm 0.34$ &  $-1.81\pm 0.48$ &  $-1.16\pm 0.33$ &  $-1.42\pm 0.34$ &  $-1.44\pm 0.61$ &  $-1.32\pm 0.45$ &  $-0.90\pm 0.43$ &  $-1.39\pm 0.14$ &  - \\
metal-rich &  $-0.24\pm 0.37$ &  $-0.67\pm 0.34$ &  $-1.17\pm 0.49$ &  $-0.58\pm 0.35$ &  $-0.54\pm 0.35$ &  $-0.54\pm 0.62$ &  $-0.35\pm 0.46$ &  $-0.11\pm 0.45$ &  $-0.54\pm 0.13$ &  - \\
\enddata
\end{deluxetable}

\begin{deluxetable}{lrrrrrrrrrrr}
\rotate
\tablewidth{0pt}
\tabletypesize{\tiny}
\tablecaption{Indices in the Lick/IDS system
\label{tab:idstab}
}
\tablehead{ Object & H$\beta$ & H$\gamma_A$ & H$\delta_A$ & H$\gamma_{\rm vaz}$ & Fe5270 & Fe5335 & \fe\ & \mgtwo\ & \mgb\ & MgFe & TiO$_{12.5}$ \\
                   & \AA      &   \AA       &   \AA       &   \AA               &  \AA   &  \AA   &  \AA &  mag    & \AA   & \AA  &  \AA   
   }
\startdata
 4 & $1.74\pm0.20$ & $1.59\pm0.26$ & $2.25\pm0.29$ & $1.34\pm0.24$ & $0.92\pm0.20$ & $1.15\pm0.26$ & $1.03\pm0.16$ & $-0.00\pm0.00$ & $1.01\pm0.18$ & $1.02\pm0.12$ & $2.1\pm7.4$ \\
 5 & $1.69\pm0.46$ & $0.55\pm0.74$ & $1.64\pm0.83$ & $1.03\pm0.81$ & $1.11\pm0.42$ & $1.16\pm0.53$ & $1.13\pm0.34$ & $0.01\pm0.01$ & $0.42\pm0.45$ & $0.69\pm0.39$ & $-24.2\pm13.6$ \\
 7 & $1.86\pm1.17$ & $3.16\pm1.22$ & $5.11\pm1.60$ & $1.46\pm1.58$ & $0.25\pm0.97$ & $0.05\pm1.36$ & $0.15\pm0.84$ & $0.03\pm0.03$ & $-0.25\pm1.11$ & $0.00\pm0.84$ & $-13.0\pm15.5$ \\
 8 & $1.61\pm0.33$ & $-4.07\pm0.37$ & $-0.39\pm0.34$ & $1.10\pm0.29$ & $1.89\pm0.27$ & $2.73\pm0.35$ & $2.31\pm0.22$ & $0.12\pm0.01$ & $2.95\pm0.25$ & $2.61\pm0.17$ & $21.9\pm6.7$ \\
 9 & $1.95\pm0.53$ & $-4.12\pm0.67$ & $-1.65\pm0.86$ & $0.17\pm0.64$ & $2.93\pm0.59$ & $2.94\pm0.60$ & $2.94\pm0.42$ & $0.17\pm0.01$ & $3.60\pm0.55$ & $3.25\pm0.34$ & $39.4\pm13.8$ \\
10 & $3.73\pm0.53$ & $-0.20\pm0.59$ & $2.20\pm0.69$ & $1.71\pm0.61$ & $1.11\pm0.54$ & $-0.68\pm0.66$ & $0.22\pm0.42$ & $0.03\pm0.01$ & $1.20\pm0.50$ & $0.51\pm0.51$ & $10.0\pm16.3$ \\
11 & $2.59\pm0.67$ & $-6.95\pm0.86$ & $-3.49\pm0.35$ & $0.37\pm0.76$ & $3.01\pm0.77$ & $0.48\pm1.11$ & $1.74\pm0.68$ & $0.14\pm0.02$ & $4.00\pm0.58$ & $2.64\pm0.55$ & $16.7\pm15.1$ \\
12 & $2.85\pm0.38$ & $-3.04\pm0.40$ & $1.66\pm0.42$ & $0.75\pm0.42$ & $0.75\pm0.39$ & $1.57\pm0.52$ & $1.16\pm0.33$ & $0.08\pm0.01$ & $1.24\pm0.51$ & $1.20\pm0.30$ & $8.6\pm7.8$ \\
13 & $2.91\pm0.82$ & $-3.85\pm1.27$ & $1.65\pm1.46$ & $0.85\pm1.31$ & $1.05\pm0.97$ & $1.90\pm1.16$ & $1.47\pm0.75$ & $0.14\pm0.02$ & $2.69\pm0.90$ & $1.99\pm0.61$ & $30.0\pm24.6$ \\
15 & $2.10\pm0.22$ & $-0.54\pm0.21$ & $2.35\pm0.21$ & $1.57\pm0.18$ & $1.70\pm0.21$ & $1.84\pm0.30$ & $1.77\pm0.18$ & $0.06\pm0.01$ & $1.52\pm0.22$ & $1.64\pm0.15$ & $28.3\pm7.0$ \\
16 & $3.31\pm0.65$ & $-1.33\pm0.90$ & $2.83\pm0.86$ & $0.87\pm0.84$ & $0.90\pm0.96$ & $1.75\pm1.25$ & $1.33\pm0.79$ & $0.05\pm0.02$ & $2.11\pm0.67$ & $1.67\pm0.57$ & $-6.4\pm24.1$ \\
17 & $1.61\pm0.58$ & $-5.00\pm0.66$ & $-3.44\pm0.94$ & $1.87\pm0.62$ & $2.48\pm0.67$ & $0.87\pm0.87$ & $1.68\pm0.55$ & $0.20\pm0.02$ & $3.77\pm0.66$ & $2.51\pm0.47$ & $34.7\pm9.7$ \\
18 & $1.41\pm0.26$ & $-6.76\pm0.26$ & $-4.81\pm0.25$ & $1.00\pm0.20$ & $3.80\pm0.26$ & $2.70\pm0.39$ & $3.25\pm0.24$ & $0.21\pm0.01$ & $4.40\pm0.26$ & $3.78\pm0.18$ & $10.9\pm8.2$ \\
19 & $1.67\pm0.38$ & $-5.36\pm0.61$ & $-1.77\pm0.77$ & $0.84\pm0.50$ & $2.30\pm0.44$ & $2.93\pm0.53$ & $2.61\pm0.35$ & $0.15\pm0.01$ & $4.32\pm0.40$ & $3.36\pm0.27$ & $32.6\pm10.4$ \\
 metal-poor & $2.39\pm0.17$ & $-0.45\pm0.23$ & $2.19\pm0.28$ & $1.25\pm0.24$ & $1.15\pm0.14$ & $1.32\pm0.24$ & $1.24\pm0.14$ & $0.05\pm0.00$ & $1.28\pm0.17$ & $1.26\pm0.11$ & $9.8\pm2.1$ \\
 metal-rich & $1.78\pm0.17$ & $-5.26\pm0.30$ & $-2.01\pm0.35$ & $0.92\pm0.23$ & $2.73\pm0.22$ & $2.26\pm0.21$ & $2.50\pm0.15$ & $0.16\pm0.01$ & $3.77\pm0.21$ & $3.07\pm0.13$ & $25.5\pm2.2$ \\
\enddata
\end{deluxetable}


\newpage

\begin{thebibliography}{}
\bibitem[Baggett, Baggett \& Anderson(1998)]{bag98}
  Baggett W. E., Baggett S. M., Anderson K. S. J., 1998, \aj, 116, 1626
\bibitem[Beasley et al.(2000)]{beas00} 
  Beasley, M.~A., Sharples, R.~M., Bridges, T.~J., Hanes, D.~A., Zepf, S.~E., 
  Ashman, K.~M., \& Geisler, D.\ 2000, \mnras, 318, 1249
\bibitem[Bridges \& Hanes(1992)]{bri92}
  Bridges T. J., Hanes D. A., 1992, \aj, 103, 800
\bibitem[Bridges et al.(1997)1997]{bri97}
  Bridges T. J., et al., 1997, \mnras, 284, 376
\bibitem[Brodie and Huchra(1990)]{bh90} 
  Brodie, J.\ P.\ and Huchra, J.\ P.\ 1990, \aj, 362, 503
\bibitem[Bruzual and Charlot(2001)]{bc2001} 
  Bruzual, G.\ A.\ and Charlot, S.\ 2001, in preparation
\bibitem[Buonanno et al.(1998)]{buo98} 
  Buonanno, R., Corsi, C.~E., Pulone, L., Fusi Pecci, F., \& 
  Bellazzini, M.\ 1998, \aap, 333, 505
\bibitem[Carney(1996)]{car96} 
  Carney, B.~W.\ 1996, \pasp, 108, 900
\bibitem[Cen(2001)]{cen01}
  Cen, R., \apj, 560, 592
\bibitem[de Freitas Pacheco \& Barbuy(1995)]{frei95} 
  de Freitas Pacheco, J.~A.~\& Barbuy, B.\ 1995, \aap, 302, 718 
\bibitem[de Vaucouleurs et al.(1991)]{devau91}
  de Vaucouleurs, G., de Vaucouleurs, A., Corwin Jr., H.\ G. et al. 1991,
  Third Reference Catalogue of Bright Galaxies, Version 3.9 (RC3)
\bibitem[Forbes, Brodie \& Larsen(2001)]{fbl01}
  Forbes, D.\ A., Brodie, J.\ P., \& Larsen, S.\ S. 2001, \apj, 556, L83
\bibitem[Forbes, Grillmair \& Smith(1997)]{for97a}
  Forbes D. A., Grillmair C. J., Smith R. C., 1997, \aj, 113, 1648
\bibitem[Fusi Pecci et al.(1993)]{fus93} 
  Fusi Pecci, F., Cacciari, C., Federici, L., \& Pasquali, A.\ 
  1993, ASP Conf.~Ser.~48: The Globular Cluster-Galaxy Connection, 410,
  editors: G.\ H. Smith and J.\ P.\ Brodie
\bibitem[Harris et al.(1984)]{har84}
  Harris W. E., Harris H. C., Harris G. L. H., 1984, \aj, 89, 216
\bibitem[Harris(1996)]{har96}
  Harris, W.\ E.\ 1996, \aj, 112, 1487
\bibitem[Huchra \& Brodie(1987)]{hb87} 
  Huchra, J.~\& Brodie, J.\ 1987, \aj, 93, 779 
\bibitem[Jones(1999)]{jon99}
  Jones, L. A. 1999, PhD Thesis, University of North Carolina
\bibitem[Kent(1988)]{ken88}
  Kent, S. M., 1988, \aj, 96, 514
\bibitem[Kissler-Patig et al.(1998)]{kis98}
  Kissler-Patig, M., Brodie, J.\ P., Schroder, L.\ et al., 1998, \aj, 115, 105
\bibitem[Kuntschner(2000)]{kun00} 
  Kuntschner, H.\ 2000, \mnras, 315, 184
\bibitem[Larsen and Brodie(2000)]{lb02}
  Larsen, S.\ S., and Brodie, J.\ P., 2002, \aj, in press
\bibitem[Larsen, Forbes \& Brodie(2001)]{lfb01}
  Larsen, S.\ S., Forbes, D.\ A., Brodie, J.\ P., 2001, \mnras, 327, 1116 
  (PaperI)
\bibitem[Maraston \& Thomas(2000)]{mt00} 
  Maraston, C.\ \& Thomas, D.\ 2000, \apj, 541, 126
\bibitem[Milone, Barbuy, \& Schiavon(2000)]{mbs00} 
  Milone, A., Barbuy, B., \& Schiavon, R.~P.\ 2000, \aj, 120, 131
\bibitem[Oke et al.(1995)]{oke95} 
  Oke, J.\ B.\, Cohen, J.\ G., Carr, M.\ et al.\ 1995, \pasp, 107, 375
\bibitem[Peletier et al.(1999)]{pel99} 
  Peletier, R.~F., Vazdekis, A., Arribas, S., del Burgo, C., 
  Garc{\' i}a-Lorenzo, B., Guti{\' e}rrez, C., Mediavilla, E., \& 
  Prada, F.\ 1999, \mnras, 310, 863
\bibitem[Rosenberg et al.(1999)]{ros99} 
  Rosenberg, A., Saviane, I., Piotto, G., \& Aparicio, A.\ 1999, \aj, 118, 2306
\bibitem[Schiavon et al.(2002)]{schia02}
  Schiavon, R.\ P., Faber, S.\ M., Rose, J.\ A., Castilho, B.\ V., 2002, \aj,
  submitted (astro-ph/0109365)
\bibitem[Schiavon et al.(2002b)]{scia02p}
  Schiavon, R.\ P., et al., 2002, in preparation
\bibitem[Stetson, Vandenberg, \& Bolte(1996)]{svb96} 
  Stetson, P.~B., Vandenberg, D.~A., \& Bolte, M.\ 1996, \pasp, 108, 560
\bibitem[Tinsley(1979)]{tin79} 
  Tinsley, B.~M.\ 1979, \apj, 229, 1046
\bibitem[Trager et al.(1998)]{trag98}
  Trager, S.~C., Worthey, G., Faber, S.~M., Burstein, D., \& Gonzalez, J.~J.\ 
  1998, \apjs, 116, 1 
\bibitem[Tripicco \& Bell(1995)]{tb95} 
  Tripicco, M.~J.~\& Bell, R.~A.\ 1995, \aj, 110, 3035
\bibitem[van den Bergh(1995)]{van95} 
  van den Bergh, S.\ 1995, \apj, 450, 27
\bibitem[Vazdekis et al.(2001)]{vaz01} 
  Vazdekis, A., Salaris, M., Arimoto, N., \& Rose, J.~A.\ 2001, \apj, 549, 274
\bibitem[Worthey(1994)]{w94}
  Worthey, G., 1994, \apjs, 95, 107
\bibitem[Worthey et al.(1994)]{wor94}
  Worthey, G., Faber, S.\ M., Gonz{\'a}lez, J.\ J., and Burstein, D.\ 1994,
  \apjs, 94, 687
\bibitem[Worthey \& Ottaviani(1997)]{wo97} 
  Worthey, G.~\& Ottaviani, D.~L.\ 1997, \apjs, 111, 377
\end{thebibliography}
\end{document}